\documentclass[aps,prd,floatfix,nofootinbib,superscriptaddress,twocolumn,showkeys]{revtex4-2}

\usepackage{amsmath}
\usepackage{amssymb}

\usepackage{makeidx}
\usepackage{amsfonts}
\usepackage[usenames,dvipsnames]{pstricks}
\usepackage{subfigure}
\usepackage{epsfig}
\usepackage{pst-grad} % For gradients
\usepackage{mathtools}
\usepackage{array}

\usepackage{allpurpose}

\newcommand\nn{{\nonumber}}

\begin{document}

\title{Deflection and gravitational lensing of null and timelike signals in general asymptotically (anti-)de Sitter spacetimes}

\author{Zixiao Li}
\thanks{These authors contributed equally to this work.}
\address{School of Physics and Technology, Wuhan University, Wuhan, 430072, China}

\author{Haotian Liu}
\thanks{These authors contributed equally to this work.}
\address{School of Physics and Technology, Wuhan University, Wuhan, 430072, China}

\author{Junji Jia}
\email[Corresponding author:~]{junjijia@whu.edu.cn}
\address{Center for Astrophysics \& MOE Key Laboratory of Artificial Micro- and Nano-structures, School of Physics and Technology, Wuhan University, Wuhan, 430072, China}

\date{\today}

\begin{abstract}
The deflection and gravitational lensing of light and massive particles in arbitrary static, spherically symmetric and  asymptotically (anti-)de Sitter spacetimes are considered in this work. We first proved that for spacetimes whose metric satisfying certain conditions, the deflection of null rays with fixed closest distance will not depend on the cosmological constant $\Lambda$, while that of timelike signals and the apparent angle in gravitational lensing still depend on $\Lambda$. 
A two-step perturbative method is then developed to compute the change of the angular coordinate and total travel time in the weak field limit. The results are quasi-series of two small quantities, with the finite distance effect of the source/detector naturally taken into account. These results are verified by applying to some known de Sitter spacetimes. Using an exact gravitational lensing equation, we solved the apparent angles of the images and time delays between them and studied the effect of $\Lambda$ on them. It is found that generally, a small positive $\Lambda$ will decrease the apparent angle of images from both sides of the lens and increase the time delay between them. The time delay between signals from the same side of the lens but with different energy however, will be decreased by $\Lambda$. 

\end{abstract}

\keywords{
deflection angle, gravitational lensing, timelike particles,  time delay, de Sitter spacetimes}

\maketitle

\section{Introduction}

Deflection of light by a massive object was one of the most striking prediction of General Relativity (GR). Its confirmation \cite{Dyson:1920cwa} not only greatly advanced the acceptance of GR among the scientific community, but also laid the foundation for the gravitational lensing (GL) phenomena. GL nowadays has become an important tool in astronomy and cosmology, due to the fact that GL observables can be connected to the properties of the source, the lens, the messengers and the spacetime the messengers went through. It was used to investigate properties of supernova \cite{Sharon:2014ija},
coevolution of galaxies and supermassive black holes (SMBHs) \cite{Peng:2006ew},
cosmological parameters \cite{Refregier:2003ct,Lewis:2006fu} and dark matter and energy \cite{Metcalf:2001ap, Hoekstra:2008db}
and to find exoplanets \cite{Mao:1991nt,MOA:2006rnt}. 

Regarding the messengers in the GL, light rays were the only practical choice. However, with the discovery of the SN 1987A supernova neutrinos \cite{Hirata:1987hu, Bionta:1987qt} and more recent blazar TXS 0506+056 neutrinos \cite{IceCube:2018dnn,IceCube:2018cha}, and the important observation of gravitational waves (GWs) after 2015 \cite{Abbott:2016blz,Abbott:2016nmj}, it was clear that neutrinos and GWs can both be the messengers in GL. These messengers are becoming particularly important partially due to the discoveries of GL of supernova \cite{Kelly:2014mwa,Goobar:2016uuf} and multi-messenger GRB-GW events \cite{Monitor:2017mdv}, and partially because the the angular and time resolution of their detectors are advancing rapidly. Meanwhile, this calls for the more careful investigation of the deflection and GL of massive timelike signals, in hope of better understanding some key problems in astronomy and particle physics, such as the supernova mechanism and neutrino properties \cite{Jia:2015zon,Pang:2018jpm,Jia:2019hih}.

In recent years, the deflection and GL of both null and timelike signals in asymptotic flat spacetimes have been considered theoretically by many authors using different methods, such as the perturbative method \cite{Jia:2015zon,Pang:2018jpm,Duan:2020tsq,Jia:2020xbc,Huang:2020trl, Liu:2020mkf} and Gauss-Bonnet theorem method \cite{Gibbons:2008rj,Werner:2012rc,Ishihara:2016vdc,Jusufi:2017lsl,Li:2019qyb,Jusufi:2017lsl}, and in different limits, including both the weak field limit (WFL) and strong field  limit \cite{Bozza:2002zj,Bozza:2007gt,Bozza:2009yw,Jia:2020xbc,Liu:2021ckg}. With the establishment of the existence of dark energy and dark matter, people are becoming increasingly interested in the deflection and GL in asymptotically (anti-)de Sitter (dS) spacetimes. Many authors studied the deflection of light rays in the Schwarzschild-dS (SdS) spacetime \cite{Ishak:2007ea,Sultana:2013ppa,Lim:2016lqv}.  Li et al. studied the strong GL of light rays in a brane world black hole (BH) \cite{Li:2015vqa}.
Panpanich et al. investigated the deflection angle of light rays in a massive gravity \cite{Panpanich:2019mll}. Se\c{c}uk et al. computed the deflection at a particular detector location in the Ressner-Nordstrom(RN)-dS-Monopole spacetime \cite{Secuk:2019svg}. Heydari-Fard studied the deflection angle in a novel 4D Gauss-Bonnet-de Sitter spacetime using the Rindler-Ishak method \cite{Heydari-Fard:2020sib}. 

However, the above investigations of deflection and GL are all done for light rays. There are still few works that considered the deflection of timelike rays in (a)dS spacetimes. Previous research either mainly concentrated on the influence of $\Lambda$ on some general features of the massive particles' trajectory, such as its perihelion shift  \cite{Cruz:2004ts,Hackmann:2008zz,Miraghaei:2010yse,Arakida:2012ya} or innermost stable circular orbits \cite{Delsate:2015ina}, or at most studied deflection of timelike rays in a particular spacetime \cite{He:2020eah}. In other words, a general method that can deal with arbitrary static and spherically symmetric (SSS) and asymptotically (a)dS spacetimes, like what Ref. \cite{Bozza:2002zj}  did to the deflection in the strong field limit, is still lacking. In this work, we attempt to establish a perturbative method that can fulfill this task. This is an extension of our previous work in the asymptotically flat spacetimes \cite{Huang:2020trl,Liu:2020wcu}. Moreover, we will also prove a result regarding the (in)dependence of the deflection angle on the cosmological constant $\Lambda$. 

This work is organized as follows. In Sec. \ref{sec:dangdefinds}, we lay out the basic metric, geodesics and definition of the deflection, and show the condition that the deflection angle is (in)dependent on the cosmological constant. In Sec. \ref{sec:dspertmethod}, the perturbative method developed in Refs. \cite{Huang:2020trl,Liu:2020wcu} is extended to the case of arbitrary SSS and asymptotically (a)dS spacetimes. Both the deflection angle and total travel time for null and timelike rays are computed, with the finite distance effect of the source and detector taken into account. These results are applied to four particular dS spacetimes in Sec. \ref{sec:spacetimecases} to verify their correctness. In Sec. \ref{sec:glcases}, the apparent angles of the GL images are solved from a new GL equation applicable to deflection angles with finite distance effect. The time delays between images from opposite sides of the lens and between images with different energies are also computed. The effects of \mclambda to these observables are carefully studied. Sec. \ref{sec:conc} concludes the work with a short discussion. Throughout the work we use the natural unit $G=c=1$ and spacetime signature $(-,+,+,+)$. 

\section{Deflection in asymptotically dS spacetimes\label{sec:dangdefinds}}

We start from the most general SSS and asymptotically dS spacetime with metric
\be
\dd s^2 = - A(r)  \dd t^2 + B(r) \dd r^2 + C(r) \lb \dd \theta^2 + \sin^2 \theta \dd \phi^2 \rb, \label{eq:metric}
\ee
where $(t,~r,~\theta,~\phi)$ are the coordinates and $A,~B,~C$ are metric functions depending on $r$ only. Without losing any generality, for SSS spacetimes we can choose the metric functions $A$ and $B$ to satisfy $A=1/B$, which is allowed (at least locally) by doing necessary changes of coordinates in \mr. For SSS and asymptotically dS spacetimes therefore we can set the metric functions to the following form
\be
A(r)=\frac{1}{B(r)}=f(r)-\frac{\Lambda r^2}{3}g(r), \label{eq:metricabform}
\ee
where $\Lambda$ is the cosmological constant. In the following, we will focus on the asymptotically dS case, i.e., assuming $\Lambda>0$. However, as one can verify, all methods and results in this work are also valid to the asymptotically anti-dS cases. In Eq. \eqref{eq:metricabform}, we will assume that $f(r)$ and $g(r)$ can be asymptotically expanded as inverse power series of $r$, 
implying that the spacetime is asymptotically flat if $\Lambda$ was zero. 

From the geodesic equations associated with metric \eqref{eq:metric}, the following equations can be obtained
\begin{align}
\frac{\dd \phi}{\dd r} = & \sqrt{\frac{B}{C}} \frac{L}{\sqrt{ \lb E^2/A - \kappa \rb C - L^2 }}, \label{eq:ddphiddr}\\
\frac{\dd t}{\dd r} = & \frac{\sqrt{BC}}{A} \frac{E}{\sqrt{ \lb E^2/A - \kappa \rb C - L^2 }}, \label{eq:ddtddr}
\end{align}
where $\kappa=0,1$ respectively for null and timelike signals and
\be
E=A(r)\frac{\dd t}{\dd \tau} ,~L=C(r)\frac{\dd \phi}{\dd \tau}
\ee
are two first integral constants, with $\tau$ being the affine parameter or proper time. In asymptotically flat spacetimes, they are interpreted respectively as the energy and angular momentum per unit mass measured by asymptotic observers. Here we will just treat them as two constants characterizing the test particle. $L$ can be related to $E$ and the minimal radius $r_0$ using the radial equation  \eqref{eq:ddtddr}, i.e., $\dd r/\dd t|_{r=r_0}=0$, to find
\be
L= \sqrt{C(r_0)\lb E^2/A(r_0) - \kappa \rb }.\label{eq:ler0rel}
\ee
In other words, among the three variables $L,~E, ~r_0$, only two of them are free. If desired, $r_0$ can be solved from Eq. \eqref{eq:ler0rel} as a function of $L,~E,~\kappa$, and $\Lambda$.
In asymptotically flat spacetimes, this $r_0$ can be related to the impact parameter $b$ of the trajectory.  Although in the asymptotically dS spacetimes, we can still define an {\it effective} impact parameter $b_{\mathrm{eff}}$ for purpose of comparison using Eq. \eqref{eq:ler0rel} for $L$
\be  b_{\mathrm{eff}}=\frac{L}{\sqrt{E^2-\kappa}}
=\frac{\sqrt{C(r_0)\lb E^2/A(r_0) - \kappa \rb }}{\sqrt{E^2-\kappa}}
\label{eq:beffdef}\ee
the interpretation of $|b|$ as the distance from the asymptotic line to the lens center is in general spoiled. 
The reason is that for asymptotically dS spacetimes with $\Lambda>0$, no signal will be able to go beyond the cosmological horizon located at $r_\mathrm{H}\approx \sqrt{\Lambda/3}$. Another important reason one should restrain from using this $b_{\mathrm{eff}}$ is that it makes the analysis of the influence of $\Lambda$ obscured because as revealed by Eq. \eqref{eq:beffdef}, $b_{\mathrm{eff}}$ itself is also $\Lambda$ dependent.

\begin{figure}[htp!]
    \centering
    \includegraphics[width=0.45\textwidth]{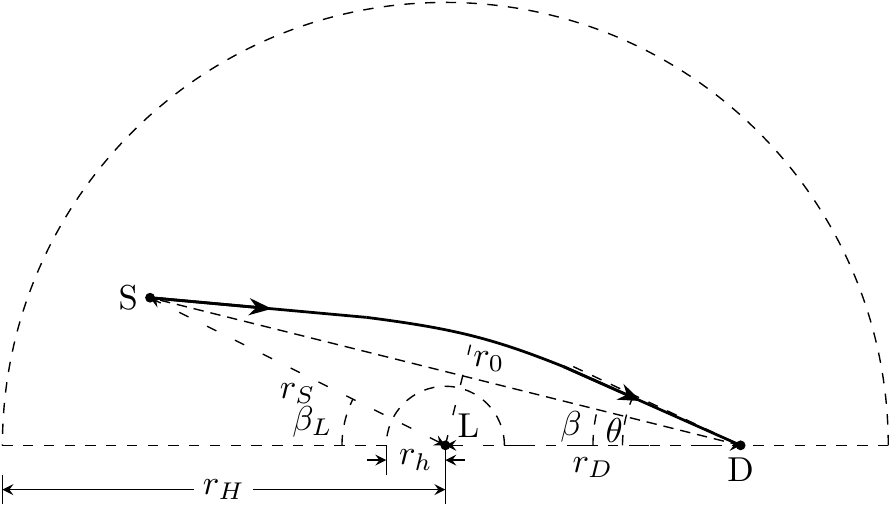}
    \caption{The deflection and GL of signals in asymptotically dS spacetimes. S, L, D represent the source, lens and detector respectively. The large dashed circle and the smaller dashed circle are respectively the cosmological horizon with radius $r_{\mathrm{H}}$ and black hole event horizon with radius $r_{\mathrm{h}}$. $\beta_L$ and $\beta$ label the angular location of the source with respect to the lens and detector respectively. The observed apparent angle is labeled as $\theta$. }
    \label{fig:sdsf1}
\end{figure}

Using Eqs. \eqref{eq:ddphiddr} and \eqref{eq:ddtddr}, the change of the angular coordinate $\Delta\phi$ and the travel time $\Delta t$ from a source at radius $r_s$ to a detector at radius $r_d$ (see Fig. \ref{fig:sdsf1}) can be computed as
\begin{align}
&\Delta \phi= \lsb \int_{r_0}^{r_s} + \int_{r_0}^{r_d} \rsb  \sqrt{\frac{B}{C}} \frac{L}{\sqrt{\lb E^2/A-\kappa\rb C  - L^2 }}\dd r, \label{eq:dphidef}\\
&\Delta t=\lsb \int_{r_0}^{r_s} + \int_{r_0}^{r_d} \rsb \frac{\sqrt{BC}}{A} \frac{E}{\sqrt{ \lb E^2/A - \kappa \rb C - L^2 }} \dd r. \label{eq:dtdef}
\end{align}
Throughout this work, we will assume that the source, lens and detector are all static so that $r_s$ and $r_d$ will not change. 

In principle, for general metric functions $A,~B$ and $C$, these two integrations can not be carried out to closed forms and therefore approximations have to be made. In this work, we will show that there exists a systematical way to perturbatively handle these integrals so that the deflection angle and travel time for both null and timelike signals can be computed to any desired accuracy for any SSS and asymptotically (a)dS spacetimes. Moreover, the method can take into account the finite distance effect of $r_s$ and $r_d$ naturally. 
Before this however, we first prove in section \ref{subsec:nullindep} that in some special cases of the spacetime and signal, the deflection $\Delta\phi$ for fixed $r_0$ is indeed independent of $\Lambda$. 

\subsection{(In)dependence of $\Delta\phi$ on $\Lambda$ for timelike(null) signals \label{subsec:nullindep}}

In this subsection, we will first show that for asymptotically dS spacetimes with 
$A(r)$ and $B(r)$ given by Eq. \eqref{eq:metricabform}, if the metric functions satisfy
\be  C(r)=h\cdot r^2g(r)\label{eq:metriccform}\ee
where $h$ is a \mclambda and $r$ independent constant, then for a null signal with fixed minimal radius $r_0$ and $L,~E$, its angular coordinate change $\Delta\phi$ as defined in Eq. \eqref{eq:dphidef} will be independent of $\Lambda$. We will further show that the apparent angle $\theta$ (see Fig. \ref{fig:sdsf1}) against the lens-detector axis in GL as measured by a detector with fixed $r_d$ however, does change as $\Lambda$ varies. Moreover, we will also show that for timelike signals, not only the apparent angle $\theta$ but also the $\Delta\phi$ itself will change as $\Lambda$ changes. 

We start from substituting Eq. \eqref{eq:ler0rel} into Eq. \eqref{eq:dphidef} and setting $\kappa=0$ for null signals, yielding
\bea
\Delta \phi
&=& \lsb \int_{r_0}^{r_s} + \int_{r_0}^{r_d} \rsb  \sqrt{\frac{A(r)B(r)C(r_0)}{C(r)\lsb A(r_0) C(r) - A(r)C(r_0) \rsb}} \dd r\nn\\
&=&\lsb \int_{r_0}^{r_s} + \int_{r_0}^{r_d} \rsb  \frac{r_0}{r}\sqrt{\frac{g(r_0)}{g(r)}}\nn\\
&&\times \sqrt{\frac{1}{h\lsb f(r_0)r^2g(r) -f(r)r_0^2g(r_0) \rsb}}\dd r, \label{eq:dphiodtnull}
\eea
where in the second step metric functions \eqref{eq:metricabform} and \eqref{eq:metriccform} are used. It is from Eq. \eqref{eq:dphiodtnull} we observe that $\Delta\phi$ does not explicitly contain or depend on $\Lambda$ for fixed $r_s,~r_d$ and $r_0$. After integration, $\Delta\phi$ becomes a function of the following form
\be
\Delta\phi=\Delta\phi(r_s,r_d,r_0,\mathrm{other~ parameters}), \label{eq:dphifunctional}
\ee
where ``other parameters'' include all the particle parameters such as its energy $E$ and spacetime parameters such as its mass $M$ and charge $Q$, but not the cosmological constant $\Lambda$.  

We emphasize that this is a general result applicable to all metrics satisfying Eq. \eqref{eq:metricabform} and \eqref{eq:metriccform}. Such metrics certainly include the familiar ones like SdS or RNdS spacetimes. Moreover, this independence holds regardless the values of $r_s,~r_d$ or $r_0$ as long as they are physically allowed. In other words, this is also applicable to source/detector near the cosmological horizon (i.e. $r_s,~r_d\to r_{\mathrm{H}}$), or to signals that experience lensing in the strong field limit (i.e. $r_0\to r_{\mathrm{photon~sphere}}$). We also point out that the independence of $\Delta\phi$ on $\Lambda$ for fixed $r_0$ was previously recognized for null rays in the SdS spacetime \cite{Lebedev:2013qoa}.

On the other hand, for timelike signals with $\kappa=1$, substituting Eq. \eqref{eq:ler0rel} and the metric functions \eqref{eq:metricabform} and \eqref{eq:metriccform} into  Eq. \eqref{eq:dphidef}, we obtain
\begin{widetext}
\begin{align}\Delta \phi
= &\lsb \int_{r_0}^{r_s} + \int_{r_0}^{r_d} \rsb \dd r \frac{r_0}{r}\frac{\sqrt{g(r_0)}}{\sqrt{g(r)}}\nn\\
&\times\sqrt{\frac{E^2- f(r_0)+\frac{\Lambda r_0^2}{3} g(r_0)}{E^2h\lsb f(r_0)g(r)r^2 -f(r)g(r_0)r_0^2\rsb-h\lsb r^2g(r)-r_0^2g(r_0)\rsb \lsb f(r)-\frac{ \Lambda r^2}{3} g(r)\rsb \lsb f(r_0)-\frac{ \Lambda r_0^2}{3} g(r_0)\rsb   }}. \label{eq:dphitimelike}
\end{align}
\end{widetext}
Here no further simplification like in the null case is possible. 
It is then clear that this $\Delta\phi$ will unavoidably depend on the value of $\Lambda$. Similarly, for the total travel time, substituting Eqs. \eqref{eq:ler0rel}, \eqref{eq:metricabform} and \eqref{eq:metriccform} and $\kappa=0$ into Eq. \eqref{eq:dtdef}, we have 
\begin{widetext}
\begin{align}
\Delta t=&
 \lsb \int_{r_0}^{r_s} + \int_{r_0}^{r_d} \rsb  \frac{ \sqrt{hr^2g(r)\lsb f(r_0)-\frac{ \Lambda r_0^2}{3} g(r_0)\rsb} }{f(r)-\frac{\Lambda r^2}{3} g(r)} \nonumber \\
  & \times \frac{E}{\sqrt{E^2h\lsb f(r_0)g(r)r^2 -f(r)r(r_0)r_0^2\rsb-h(r^2g(r)-r_0^2g(r_0))\lsb f(r)-\frac{ \Lambda r^2}{3} g(r)\rsb \lsb f(r_0)-\frac{ \Lambda r_0^2}{3} g(r_0)\rsb   }} \dd r. 
\label{eq:totaltsub}
\end{align}
\end{widetext}
It is also clear that the total travel time and consequently the time delay between the images will all depend on $\Lambda$, even for light rays. For timelike rays, we can verify similarly to $\Delta\phi$ in Eq. \eqref{eq:dphitimelike} that the dependence of $\Delta t$ on $\Lambda$ is even more apparent. 

\subsection{Dependence of the apparent angle on $\Lambda$}

The apparent angle of the light ray observed by a static observer against the observer-lens axis however is not $\Delta\phi$ but $\theta$, as illustrated in Fig. \ref{fig:sdsf1}. As pointed out by Bhadra et al. \cite{Bhadra:2010jr}, $\theta$ is indeed dependent on $\Lambda$ for light rays in the SdS spacetime. In this part, we will briefly demonstrate that this is generally true for other asymptotically dS spacetimes too. The angle $\theta$ in the SSS spacetimes described by metric \eqref{eq:metric} can be computed using the projection of the four velocity of the signal 
\be
\frac{\dd x^{\mu}_{\mathrm{sig}}}{\dd \tau}=\lsb \frac{E}{A},~\sqrt{E^2-\frac{L^2A}{C}-\frac{\kappa}{B}},~\frac{L}{C}\rsb
\ee
onto that of the static observer \cite{Hobson:2006se,Huang:2020trl}
\be \frac{\dd x^\mu_{\mathrm{obs}}}{\dd \tau}=\lsb\frac{1}{\sqrt{A(r_d)}},~0,~0,~0 \rsb.\ee  It is straightforward to show that the resultant apparent angle $\theta$ for metric \eqref{eq:metric} is 
\bea
\theta&=&\arcsin\lb \frac{L}{\sqrt{E^2-\kappa A(r_d)}}\sqrt{\frac{A(r_d)}{C(r_d)}}\rb\nn\\
&=&\arcsin\lb \sqrt{\frac{E^2-\kappa A(r_0)}{E^2-\kappa A(r_d)}}\sqrt{\frac{A(r_d)C(r_0)}{A(r_0)C(r_d)}}\rb,
\label{eq:aasol}\eea
where in the second step Eq. \eqref{eq:ler0rel} is substituted for $L$.

The apparent angle \eqref{eq:aasol} at a freely varying $b_{\mathrm{eff}}$ and a particular observer location $r_b$ determined by the alignment of the source, lens and observer (see Eq. (15) of Ref. \cite{He:2020eah} for its value in SdS spacetime), was previously computed for light rays and timelike signals in SdS spacetime in Refs. \cite{Ishak:2007ea,Sultana:2013ppa,Lim:2016lqv} and \cite{He:2020eah} respectively. Although in these works, $\theta $ was called the ``deflection angle'' and this choice of $r_b$ is not well justified \cite{Bhadra:2010jr}, nevertheless we plugged $r_d=r_b$ into Eq. \eqref{eq:aasol} and verified that Eq. \eqref{eq:aasol} will correctly yield the corresponding values in these cases, i.e., Eq. (15) of Ref. \cite{Ishak:2007ea}, Eq. (17) of Ref. \cite{Sultana:2013ppa} (with $a=0$), Eq. (31) of Ref. \cite{Lim:2016lqv} and Eq. (18) of Ref. \cite{He:2020eah}. 

For light signal with $\kappa=0$, Eq. \eqref{eq:aasol} simplifies to
\be
\theta=\arcsin\lb \sqrt{\frac{A(r_d)C(r_0)}{A(r_0)C(r_d)}}\rb.
\label{eq:aasollight}\ee
In a GL configuration, if $\Lambda$ is changed while holding source and detector static, i.e., $r_s,~r_d$ and $\beta_L$ in Fig. \ref{fig:sdsf1} unchanged, then it is clear from Fig. \ref{fig:sdsf1} that $\Delta\phi=\pi\pm \beta_L$ is also kept as a constant. From Eq. \eqref{eq:dphifunctional}, this in turn means that for null signal, as $\Lambda$ varies, $r_0$ should not change either. Then immediately using Eq. \eqref{eq:aasollight} we see that $\theta$ will vary since all variables in the metric function $A(r_0)$ or $A(r_d)$ are unchanged except $\Lambda$.

For timelike signal, we have shown previously that unlike null signal, its $\Delta\phi$ will change with the variation of $\Lambda$. For its apparent angle $\theta$, its also seen from Eq. \eqref{eq:aasol} by setting $\kappa=1$ that it will depend on the parameter $\Lambda$ through $A(r_0)$ and $A(r_d)$, just as in the case of null signal.
Finally we also note that if the spacetime is expanding, as in $\Lambda$CDM standard cosmology, an observer can not be truly static and therefore its motion should also be taken into account when considering the effect of \mclambda on observables such as the apparent angle \cite{Butcher:2016yrs}. In those cases, which is beyond the scope of the current work, the conclusion about the effect of \mclambda might be different from the SSS case considered here. 

\section{Perturbative deflection in the WFL \label{sec:dspertmethod}}

In this section, we will resume computing $\Delta\phi$ and $\Delta t$ according to Eqs. \eqref{eq:dphidef} and \eqref{eq:dtdef} in the WFL for asymptotically dS spacetimes.
In this limit, we can safely assume that \be
M\ll r_0\ll \{ r_s, ~r_d\}< r_{\mathrm{H}} \approx \sqrt{3/\Lambda},
\label{eq:rrels}
\ee 
where $M$ is the mass of the spacetime.

\subsection{The change of angular coordinate $\Delta\phi$ \label{subsec:dang}}
Starting from definition \eqref{eq:dphidef}, our goal here is to find a perturbative way to systematically expand the integrand, which should then allow us to carry out the integration and find an approximation of $\Delta\phi$. The main difficulty in perturbative expansion in the asymptotically dS spacetime in the WFL, comparing to that in asymptotically flat spacetime, is that we can not straightforwardly carry out the large $r_0/M$ (or $b_{\mathrm{eff}}/M$) expansion because the source or observer can not be at a radius larger than the cosmological horizon $r_{\mathrm{H}}$. 
Therefore in this work, we use a trick to hold $r_0/r_{\mathrm{H}}$ small while doing the large $r_0$ expansion. We first define a small variable $\epsilon$ 
\be 
\epsilon\equiv r_0\sqrt{\Lambda}\sim r_0/r_{\mathrm{H}}\mathrm{~so~ that~}r_0=\frac{\epsilon}{\sqrt{\Lambda}}. \label{eq:r0lrel}
\ee 
Then we can make a change of variable in Eq. \eqref{eq:dphidef} from $r$ to the new integral variable $u$
\be 
r\to u=\frac{r}{r_0}~~\text{or equivalently}~~ r=ur_0=\frac{u\epsilon}{\sqrt{\Lambda}}. \label{eq:cov} \ee 
Using this, the integral limits of Eq. \eqref{eq:dphidef} and the Jacobian become
\bea
&& r_0 \to 1, ~
r_{s,d} \to \frac{r_{s,d}\sqrt{\Lambda}}{\epsilon}\equiv u_{s,d}, \\
&& \dd r \to \frac{\epsilon}{\sqrt{\Lambda}}\dd u. 
\eea
Substituting Eq. \eqref{eq:ler0rel} into Eq. \eqref{eq:dphidef} and using the change of variables \eqref{eq:cov} for all $r$, $\Delta\phi$ becomes
\begin{widetext}
\be
\Delta \phi= \lsb \int_1^{u_s} + \int_1^{u_d} \rsb   \sqrt{\frac{B\lb \frac{u\epsilon}{\sqrt{\Lambda}}\rb}{C\lb \frac{u\epsilon}{\sqrt{\Lambda}}\rb}} \frac{\sqrt{\lsb \frac{E^2}{A\lb\frac{\epsilon}{\sqrt{\Lambda}}\rb} - \kappa \rsb C\lb \frac{\epsilon}{\sqrt{\Lambda}}\rb}}{\sqrt{ \lsb \frac{E^2}{A\lb\frac{u\epsilon}{\sqrt{\Lambda}}\rb} - \kappa \rsb C\lb \frac{u\epsilon}{\sqrt{\Lambda}}\rb - \lsb \frac{E^2}{A\lb\frac{\epsilon}{\sqrt{\Lambda}}\rb} - \kappa \rsb C\lb \frac{\epsilon}{\sqrt{\Lambda}}\rb }}
\frac{\epsilon}{\sqrt{\Lambda}}\dd u.
\label{eq:dphires}
\ee
\end{widetext}

Since it is assumed that the spacetime is asymptotically flat if $\Lambda$ were zero, we can set the $f(r),~g(r)$ and  $C(r)$ in Eq. \eqref{eq:metricabform} to the following expansion form at large $r$
\be 
f(r)=\sum_{n=0}\frac{f_n}{r^n},~
g(r)=\sum_{n=0}\frac{g_n}{r^n},~
C(r)=r^2\sum_{n=0}\frac{c_n}{r^n}.
\label{eq:fcexp}\ee
Without losing any generality, we can always set $f_0=g_0=c_0=1$ by a scaling of coordinates $t,~r$ and constant $\Lambda$. Moreover, we can always identify $f_1=-2M$ where $M$ is the ADM mass of the spacetime. 

To perturbatively compute $\Delta\phi$ in Eq. \eqref{eq:dphires}, we propose a two-step perturbative method, to expand the integrand for small $\Lambda$ first and then for small $\epsilon$. According to Eq. \eqref{eq:r0lrel}, expanding around small $\Lambda$ is essentially equivalent to a large $r_0$ expansion, while the subsequent small $\epsilon$ expansion makes sure that $r_0/r_{\mathrm{H}}$ is still kept small. We emphasize that in order for the method to work and to have a proper interpretation of the result, it is important to carry out the expansion in the above order.  Substituting Eqs. \eqref{eq:metricabform} and \eqref{eq:fcexp} into \eqref{eq:dphires} and carrying out the said expansion,  $\Delta\phi$ is found to be
\begin{align}
\Delta\phi=&\lsb \int_{1}^{u_s} + \int_{1}^{u_d} \rsb \sum_{n=0}^{\infty} \frac{\Lambda^{\frac{n}{2}} }{\epsilon^n (u+1)^n } \nn\\
&\times\lcb \sum_{m=0}^{\infty} \epsilon^{2m} \lsb \sum_{k=1-n-2\delta_{m0}}^{n+2m-1} p_{n,m,k} \frac{u^k}{\sqrt{u^2-1}} \rsb \rcb \dd u, \label{eq:dphisep}
\end{align}
where $p_{n,m,k}$ are the expansion coefficients determined by the metric function coefficients and $\delta_{m0}$ is the Kronecker delta function. The first several $p_{n,m,k}$ are given by
\begin{subequations}
\label{eq:pfirstfew}
\begin{align}
&p_{0,0,-1}=1, \\
&p_{0,1,1}=\frac{-\kappa}{6(E^2-\kappa)} , \\
&p_{1,0,-2}=p_{1,0,-1}=-\frac{f_1 +c_1}{2}, \\
&p_{1,0,0}=-\frac{E^2(f_1-c_1)+\kappa  c_1 }{2  (E^2-\kappa) },\\
&p_{1,1,0}=\frac{E^4 \left(2 g_1-2 c_1\right)}{12 \left(E^2-\kappa \right)^2} \nn \\
&+\frac{\kappa  \left[E^2 \left(3 c_1+5 f_1-4 g_1\right)-c_1-3 f_1+2 g_1\right]}{12 \left(E^2-\kappa \right)^2},\\
&p_{1,1,1}=\frac{\kappa  \left[E^2 \left(c_1+f_1-2 g_1\right)-c_1-3 f_1+2 g_1\right]}{12 \left(E^2-\kappa \right)^2},\\
&p_{1,1,2}=\frac{\kappa  \left[E^2 \left(f_1-c_1\right)+c_1\right]}{12 \left(E^2-\kappa \right)^2}.
\end{align}
\end{subequations}
The higher order ones can be obtained easily too but are more lengthy and therefore not presented here except in some specific spacetimes in Sec. \ref{sec:spacetimecases}. Note that since the dimension of $f_n,~g_n$ and $c_n$ are $M^n$ where $M$ is the mass of the spacetime, from Eq. \eqref{eq:pfirstfew} we see that the dimension of $p_{n,m,k}$ is also $M^n$. One would expect that if $f_n,~g_n$ and $c_n$ are comparable to or smaller than $M^n$, $p_{n,m,k}$ will also be smaller than or at most comparable to $M^n$, . 

Now the computation of $\Delta\phi$ in Eq. \eqref{eq:dphisep} relies on the integrability of the following integrals \bea 
I_{n,k}(u_{s,d})&\equiv &\int_1^{u_{s,d}}
\frac{u^k}{(u+1)^n\sqrt{u^2-1}}\dd u\\
&=&
\int_{0}^{\alpha_{s,d}} \frac{\cos^{n-k-1}\alpha }{\lb \cos\alpha +1 \rb^n} \dd\alpha, \label{eq:ttintfuncu}
\eea
where in the second step a change of variables $u=\sec\alpha$ is used, and $ \alpha_{s,d}\equiv \sec^{-1}(u_{s,d})$ are defined. We can prove that the integral \eqref{eq:ttintfuncu}
can always be carried out for general $n$ and $k$. Their results are elementary functions given in Eq. \eqref{eq:inkres}. Later, to study the finite distance effect of $r_{s,d}$ and to solve the lensing equations, we can also expand $I_{n,k}$'s in the large $u_{s,d}$ limit, whose leading order results are presented in Eqs. \eqref{eq:inkgeneralexp} and \eqref{eq:i1nexp}. 
Substituting $I_{n,k}$ back into Eq. \eqref{eq:dphisep} and replacing $\epsilon$ by $r_0$ using \eqref{eq:r0lrel}, we have finally
\be
\Delta\phi  =\sum_{j=s,d}
\sum_{n=0,m=0}^{\infty} \Delta\phi_{n,m} \label{eq:alphaintres}
\ee
where the contribution from order $1/r_0^n$ and $(r_0^2\Lambda)^m$ is
\be 
\Delta\phi_{n,m}=\frac{(r_0^2\Lambda)^m}{r_0^n } \sum_{k=1-n-2\delta_{m0}}^{n+2m-1} p_{n,m,k} I_{n,k}\lb \frac{r_j}{r_0}\rb.
\ee

A few comments are in order for  result \eqref{eq:alphaintres}. Firstly, this result is a quasi-power series of two  variables, namely $1/r_0$ and $r_0^2\Lambda$. The simultaneous appearance of $r_0^2$ and $\Lambda$ in $(r_0^2\Lambda)^m$ makes sure that the large $r_0$ limit will not cause divergence in $\Delta\phi$ since $r_0<r_{\mathrm{H}}\sim \sqrt{3/\Lambda}$. Secondly, note from the expression \eqref{eq:pfirstfew} that for null signal and if $c_n=h\cdot g_n$ for all $n\geq 0$ ($h$ was fixed to 1 when setting $c_0=g_0=1$), all $p_{n,m,k}$ with $m\geq 1$ indeed will reduce to zero. Since the $m\geq 1$ terms in $\Delta\phi$ are exactly the terms containing $\Lambda$, the above is just a manifestation of the independence of $\Delta\phi$ on $\Lambda$ for null signals in spacetime satisfying Eq. \eqref{eq:metriccform}, as proven in Sec. \ref{subsec:nullindep}. It is also clear that even for timelike rays with $\kappa=1$, the $m\geq 1$ terms are generally suppressed in the ultra-relativistic limit $(E\to\infty)$ by the $1/E^{2m}$ factor and therefore the effect of $\Lambda$ will be less obvious. 

Further noticing the order estimation 
\eqref{eq:inkgeneralexp}, for the $k$ summation in Eq. \eqref{eq:alphaintres} we see that for each fixed $m$ and $n$, the term with highest index $k=n+2m-1$ will dominate. Therefore keeping only these terms, $\Delta\phi$ in Eq. \eqref{eq:alphaintres} should roughly be
\begin{align}
\Delta\phi  \approx\sum_{j=s,d}&\lsb
\sum_{n=0}^{\infty} \frac{L_{n,n-1} }{r_0^n } p_{n,0,n-1} \right.\nn\\
+\sum_{n=0,m=1}^{\infty} &\left. \frac{1}{2m-1} \frac{1 }{r_0^n } (r_0^2\Lambda)^m   p_{n,m,n+2m-1} \lb \frac{r_j}{r_0}\rb^{2m-1} \rsb \nn\\
&=\sum_{j=s,d}\lsb 
\sum_{n=0}^{\infty} \frac{L_{n,n-1}}{r_0^n } p_{n,0,n-1} \right.\nn\\
&\left.+\frac{r_0}{r_{s,d}}\sum_{n=0,m=1}^{\infty} \frac{1}{2m-1}\frac{1 }{r_0^n} (r_{s,d}^2\Lambda)^m   p_{n,m,n+2m-1}\rsb,
\label{eq:dphiorderest}
\end{align}
where $L_{n,n-1}~(n\geq 0)$ are $r_0$ and $\Lambda$ independent constants given in Eq. \eqref{eq:inkgeneralexp}. 
Since the coefficients $p_{n,m,k}$ are of dimension $M^n$, if the coefficients $f_n, g_n$ and $c_n$ are not much larger than $M^n$, then as an order estimation we should have
\bea \mathcal{O}\lb \Delta\phi \rb
&=&
\sum_{n=0}^{\infty} \lb \frac{M}{r_0}\rb^n \cdot  \mathcal{O}(1) \nn\\
&&+\frac{r_0}{r_{s,d}}\sum_{n=0,m=1}^{\infty} \frac{M^n }{r_0^n} (r_{s,d}^2\Lambda)^m \cdot  \mathcal{O}(1).
\eea
Clearly, $\Delta\phi$ is a quasi-series of product of two series of two small parameters, $M/r_0$ and $r_{s,d}^2\Lambda$ respectively. The dominance of the terms depends on the detailed numerical values of these parameters. 
In Sec. \ref{sec:glcases}, we will consider two limiting cases: the first being $M/r_0$ as large as possible and  $r_{s,d}^2\Lambda$ is small, and the second case being $r_{s,d}^2\Lambda$ as large as possible and $M/r_0$ small.

\subsection{The total travel time}

For the total travel time \eqref{eq:dtdef}, we can carry out a procedure similar to steps from Eq. \eqref{eq:rrels} to Eq. \eqref{eq:alphaintres}. After the change of variables \eqref{eq:cov} and expansion of $\Lambda$ and then $\epsilon$, $\Delta t$ becomes
\begin{align}
\Delta t=&\lsb \int_{1}^{u_s} + \int_{1}^{u_d} \rsb \sum_{n=0}^{\infty} \frac{\Lambda^{\frac{n-1}{2}} }{\epsilon^{n-1} (u+1)^n } \nn\\
&\times \lcb \sum_{m=0}^{\infty} \epsilon^{2m} \lsb \sum_{k=1-n}^{\max[n-2,0]+2m+1} q_{n,m,k} \frac{u^k}{\sqrt{u^2-1}} \rsb \rcb \dd u, \label{eq:genttintform2}
\end{align}
where $q_{n,m,k}$ are related to the coefficients of the metric functions. The first few of them are
\begin{subequations}
\label{eq:qfirstfew}
\begin{align}
q_{0,0,1} = & \frac{E}{ \sqrt{E^2- \kappa }}, \\
q_{0,1,1} = & - \frac{E^3 }{6 \lb E^2-\kappa \rb^{\frac{3}{2} } }, \\
q_{0,1,2} = & 0, \\
q_{0,1,3} = & \frac{E \lb 2 E^2 - 3 \kappa \rb}{6 \lb E^2-\kappa  \rb^{\frac{3}{2} } }, \\
q_{1,0,0} = & \frac{E\lb c_1 - 3 f_1 \rb }{2  \lb E^2-\kappa \rb^{\frac{1}{2} }}, \\
q_{1,0,1} = & -\frac{Ef_1 \lb 2 E^2 - 3 \kappa \rb}{2 \lb E^2-\kappa \rb^{\frac{3}{2} }}, 
\end{align}
\end{subequations}
and higher order ones can be similarly obtained without any difficulty. 
The qualitative difference of Eq. \eqref{eq:genttintform2} from Eq. \eqref{eq:dphisep} is that the beginning order of $\sqrt{\Lambda}/\epsilon$ is from $-1$. This is expected because $\Delta t$ should carry a dimension of $[\sqrt{\Lambda}/\epsilon]^{-1}=[r_0]$. 

Eq. \eqref{eq:genttintform2} can also be integrated, yielding the result
\be
\Delta t=\sum_{j=s,d}
\sum_{n=0,m=0}^{\infty} \Delta t_{n,m}, \label{eq:tintres}
\ee
where 
\be
\Delta t_{n,m}= \frac{ (r_0^2\Lambda)^m}{r_0^{n-1} } \sum_{k=1-n}^{\max[n-2,0]+2m+1} q_{n,m,k} I_{n,k}\lb \frac{r_j}{r_0}\rb , \label{eq:dtnmres}
\ee
with $I_{n,k}$ given by Eq. \eqref{eq:inkres}.
In particular, using the first few $I_{n,k}$ in Eq. \eqref{eqs:inks}, to the order $\mathcal{O} \lb r_0^3 \Lambda \rb$, the total time is
\begin{widetext}
\begin{align}
\Delta t
 = & \sum_{i=s,d} \lcb q_{0,0,1} r_0 \sqrt{\lb\frac{r_i}{r_0}\rb^2-1} + r_0^3 \Lambda \lsb q_{0,1,1} \sqrt{\lb\frac{r_i}{r_0}\rb^2-1} \right.  
 + \frac{q_{0,1,3}}{3} \sqrt{\lb\frac{r_i}{r_0}\rb^2 -1} \lb \lb\frac{r_i}{r_0}\rb^2+2\rb \rsb  + q_{1,0,0} \sqrt{\frac{r_i/r_0-1}{r_i/r_0+1} } \nonumber \\
 & \left. + q_{1,0,1} \lsb \ln \lb \sqrt{\lb\frac{r_i}{r_0}\rb^2-1} + \frac{r_i}{r_0} \rb - \sqrt{\frac{r_i/r_0-1}{r_i/r_0+1} } \rsb \rcb  
 + \mathcal{O} \lb \frac{M}{r_0}, r_0^5 \Lambda^2 \rb, \nn \\
 = & \sum_{i=s,d} \frac{E}{ \sqrt{E^2- \kappa }} \lcb  r_0 \lb \frac{r_i}{r_0} - \frac{r_0}{2r_i} \rb + \frac{r_0^3 \Lambda}{6 \lb E^2-\kappa  \rb }   \lsb \frac{ \lb 2 E^2 - 3\kappa \rb r_i^3}{3 r_0^3}-\frac{3 \kappa r_i}{2 r_0} - \frac{ \lb 2 E^2 - 9 \kappa \rb r_0 }{8 r_i} \rsb \right.  \nonumber \\
  & \left.  + \frac{\lb E^2-\kappa \rb c_1 - E^2 f_1}{2 \lb E^2-\kappa \rb} \lb 1 - \frac{r_0}{r_i} \rb -\frac{f_1 \lb 2 E^2 - 3 \kappa \rb}{2 \lb E^2-\kappa \rb}  \ln \lb \frac{2r_i}{r_0} \rb  \rcb + \mathcal{O} \lb \frac{M}{r_0}, r_0^5 \Lambda^2, \frac{r_0^2}{r_i^2} \rb ,  \label{eq:totaltfirstfew}
\end{align}
\end{widetext}
where in the second step the $q_{n,m,k}$ given in Eq. \eqref{eq:qfirstfew} and the expansion of $I_{n,k}$ in Eq. \eqref{eq:i1nexp} are used. 
The first, second and last two terms in the curl brackets are respectively due to the $\Delta t_{0,0},~\Delta t_{0,1}$ and $\Delta t_{1,0}$ terms in Eq. \eqref{eq:tintres}.
Later in Sec. \ref{subsec:td}, we will show that the total time truncated to the above order will determine the leading three orders of the time delay between the GL images. 
One particularly interesting limit of $\Delta t$ is its ultra-relativistic limit. Expanding Eq. \eqref{eq:totaltfirstfew} in the infinite $E$ limit, we obtain 
\begin{align} 
\Delta t
 \approx & \sum_{i=s,d} \lcb r_i-\frac{r_0^2}{2r_i} + \frac{\Lambda}{3} \lb \frac{r_i^3}{3}-\frac{r_0^4}{8r_i} \rb  \right. \nonumber \\
  & + \frac{c_1-a_1}{2}\lb 1-\frac{r_0}{r_i} + \frac{r_0^2}{2 r_i^2} \rb - a_1 \ln \frac{8r_i}{r_0} \nn\\
  &+ \frac{1}{E^2} \lsb \frac{r_i}{2} - \frac{r_0^2}{4 r_i} + \frac{\Lambda}{4} \lb -r_0^2 r_i+ \frac{r_0^4}{2r_i} \rb\right. \nonumber \\
  &\left. \left.  + \frac{c_1-3a_1}{4} 
\lb 1-\frac{r_0}{r_i} + \frac{r_0^2}{r_i^2} \rb  \rsb \rcb  \nonumber \\
  & + \mathcal{O} \lb \frac{M}{r_0}, r_0^5 \Lambda^2, \frac{r_0^2}{r_i^2}, \frac{1}{E^4} \rb  .\label{eq:tdlargee}
\end{align}
The first two lines represent the leading orders of the travel time of null rays and the third and fourth lines are the relativistic corrections. They will be useful when considering the time delay between different kinds of signals in Sec. \ref{subsec:td}. 

\section{Application to asymptotically dS spacetimes \label{sec:spacetimecases}}

To check the validity of the perturbative method presented in Sec. \ref{sec:dspertmethod}, especially the change of the angular coordinate \eqref{eq:alphaintres}, in this section we will compute $\Delta \phi$ in a few known asymptotically dS spacetimes. 

\subsection{Schwarzschild-dS Spacetime\label{subsec:sds}}

We first consider the SdS spacetime case, which has a metric of Eq. \eqref{eq:metric} with \eqref{eq:metricabform} and
\be 
f(r)=1-\frac{2M}{r},~g(r)=1,~C(r)=r^2. \ee
From this, reading off the coefficients $f_i,~g_i$ and $c_i$ in Eq. \eqref{eq:fcexp} and substituting into Eq. \eqref{eq:pfirstfew} and further into \eqref{eq:alphaintres}, we immediately obtain $\Delta\phi$ in SdS spacetime for both null and timelike signals, with finite distance effect taken into account. To the order $\mathcal{O}(M/r_0)^2$ and $\mathcal{O}(r_0^2\Lambda)^1$, we have
\begin{align} 
\Delta\phi_{\mathrm{S}}=&
\sum_{j=s,d} \lsb \Delta\phi_{\mathrm{S},0,0} +
\Delta\phi_{\mathrm{S},0,1} +\Delta\phi_{\mathrm{S},1,0} +\Delta\phi_{\mathrm{S},1,1}\right.\nn\\
&\left.+\Delta\phi_{\mathrm{S},2,0} \rsb+\mathcal{O}\lsb \lb \frac{M}{r_0}\rb^3,~(r_0^2\Lambda)^2\rsb
\label{eq:dphiressch}
\end{align}
where the contributions from each order are
\begin{subequations}
\label{eq:dphisnm}
\begin{align}
\Delta\phi_{\mathrm{S},0,0}=& I_{0,-1}, \label{eq:dphis00}\\
\Delta\phi_{\mathrm{S},0,1}=&-(r_0^2 \Lambda)\frac{\kappa}{6(E^2-\kappa)} I_{0,1}, \label{eq:dphis01}\\
\Delta\phi_{\mathrm{S},1,0}=& \frac{M}{r_0} \lb  I_{1,-2}+ I_{1,-1}+\frac{ E^2}{E^2-\kappa}I_{1,0} \rb, \label{eq:dphis10} \\
\Delta\phi_{\mathrm{S},1,1}=&\frac{M}{r_0}(r_0^2 \Lambda)\nn\\
&\times\frac{\kappa \lsb (3 -5 E^2)I_{1,0}+(3 -E^2)I_{1,1}-E^2I_{1,2}\rsb }{6(E^2-\kappa)^2},\\
\Delta\phi_{\mathrm{S},2,0}=&\frac{M^2}{r_0^2}
 \lsb \frac32  I_{2,-3} +3  I_{2,-2}+\frac{3  \left(3 E^2-\kappa \right)}{2(E^2-\kappa) } I_{2,-1}\right.\nn\\
&\left.\left. + \frac{ E^2 \left(3 E^2-5 \kappa \right)}{(E^2-\kappa )^2} I_{2,0}+\frac{E^2 \left(3 E^2-4 \kappa \right)}{2(E^2-\kappa )^2} I_{2,1}\rsb \rcb. \label{eq:dphis20}
\end{align}
\end{subequations}
Note the higher orders are obtained but not presented here due to their length.
To see the finite distance effect of the source and detectors more clearly, using the expansions \eqref{eq:i1nexp} for $I_{n,k}$,  the $\Delta\phi_{\mathrm{S},n,m}$ become to the leading orders of $r_0/r_i~(i=s,d)$
\begin{subequations}
\label{eq:dphiresschinf}
\begin{align}
\Delta\phi^\prime_{\mathrm{S},0,0}=&  \frac{\pi}{2}-\frac{r_0}{r_j}, \label{eq:dphiresschinf00}\\
\Delta\phi^\prime_{\mathrm{S},0,1}=&-(r_0^2 \Lambda)\frac{\kappa}{6(E^2-\kappa)} \left(\frac{r_j}{r_0}-\frac{r_0}{2r_j}\right), \label{eq:dphiresschinf01}\\
\Delta\phi^\prime_{\mathrm{S},1,0}=&\frac{M}{r_0}  \lsb 1+\frac{ E^2}{E^2-\kappa}\lb 1-\frac{r_0}{r_j}\rb \rsb , \label{eq:dphiresschinf10} \\
\Delta\phi^\prime_{\mathrm{S},1,1}=&\frac{M}{r_0} (r_0^2 \Lambda)\nn\\
&\times\frac{\kappa  \lcb2 r_0 r_j \lsb 3 \kappa  \ln \left(\frac{2 r_j}{r_0}\right)-5 E^2\rsb-2 E^2 r_j^2\rcb}{12  \left(E^2-\kappa \right)^2 r_0r_j}, \label{eq:dphiresschinf11}\\
\Delta\phi^\prime_{\mathrm{S},2,0}=&\frac{M^2}{r_0^2}
\frac{\lsb E^4\lb 15\pi-16\rb +2E^2\kappa \lb 4-9\pi\rb + 3 \pi\kappa \rsb }{ 8 (E^2-\kappa )^2}.\label{eq:dphiresschinf20}
\end{align}
\end{subequations}
Here we kept the result to different orders of $(r_0/r_j)$ in different $\Delta\phi_{\mathrm{S},n,m}$ because when they blend into $\Delta\phi_\mathrm{S}$, a combined second order of the small quantities $M/r_0,~ r_0^2\Lambda$ and $r_0/r_j$ can be achieved. Setting $\Lambda=0$ in Eqs. \eqref{eq:dphisnm} and \eqref{eq:dphiresschinf} reduces them to the Schwarzschild results obtained previously for both timelike and null rays \cite{Jia:2020xbc}. On the other hand, setting $\kappa=0$, the $(r_0^2\Lambda)^n~(n\geq 1)$ terms automatically disappear from $\Delta\phi_\mathrm{S}$ and the result becomes that of null rays \cite{Jia:2020xbc}. If one is interested in the deflection of null rays expressed in terms of $b_\mathrm{eff}$, then replacing $r_0$ by $b_\mathrm{eff}$ using \eqref{eq:beffdef} and setting $\kappa=0$ in Eq. \eqref{eq:dphiresschinf}, the Eq. (5) of Ref. \cite{Sereno:2007rm} can be recovered. On the other hand, if one is interested in the null deflection from a source to an observer both located asymptotically at the cosmological horizon, then setting $\kappa=0$ and  $r_s=r_d=r_\mathrm{H}$ in Eq. \eqref{eq:dphiresschinf}, the Eq. (55) of Ref. \cite{Batic:2014loa} is obtained. 

\begin{figure}[htp!]
    \centering
\includegraphics[width=0.45\textwidth]{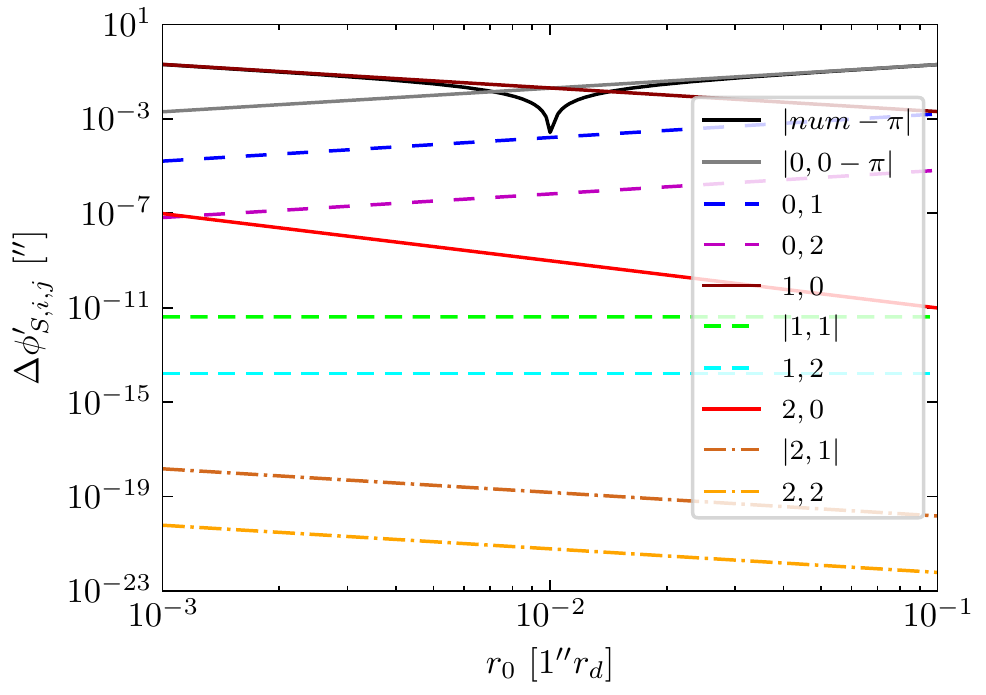}\\
    (a)\\
    \includegraphics[width=0.45\textwidth]{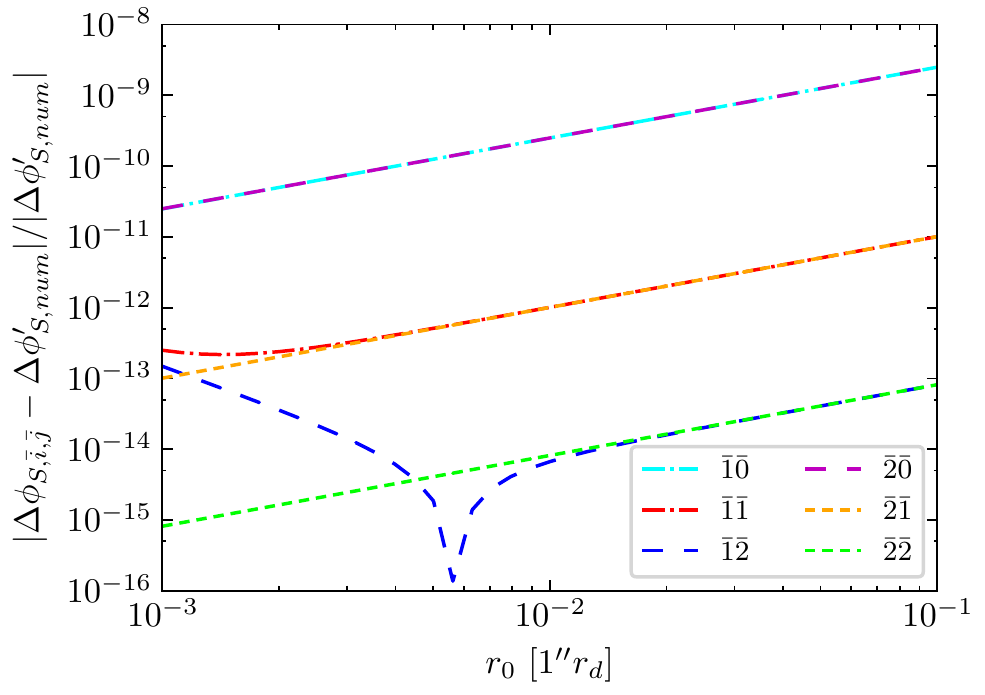}\\
    (b)
    \caption{$\Delta\phi_{\mathrm{S}}$ using Eq. \eqref{eq:dphiresschinf}. The brown and red solid curves denote respectively the result \eqref{eq:dphiresschinf} truncated to the $r_0^{-1}$ and $r_0^{-2}$ orders without $\Lambda^1$ terms, and the blue dashed curved denotes the result including the $\Lambda^1$ terms. \label{fig:sdsplot}}
\end{figure}
    
To verify the correctness of these results, in Fig. \ref{fig:sdsplot} (a) we plot the contributions from different $\Delta\phi_{\mathrm{S},n,m}^\prime$ in Eq. \eqref{eq:dphiresschinf}.
In cosmology, it is known through Planck 2018 observation that the cosmological constant is quite small $\Lambda_c\leq 1.06\times 10^{-52}$ [m$^{-2}$] \cite{Aghanim:2018eyx}. To see the effect of $\Lambda$ clearly among the contributions from all terms, we have to use very large $r_0,~r_{s,d}$ and relatively small $M$. Therefore we chose a toy dwarf galaxy whose distance is $r_d=13.4$ billion ly \cite{Jiang:2020rdl} with a mass $M=10^6M_\odot$  as the lens. For energy $E$, as pointed out in Sec. \ref{sec:dspertmethod}, the infinite \mce limit will force $\Delta\phi$ to approach its value of null rays and become independent on $\Lambda$ in the SdS case. Therefore in the plots, we will only set \mce to a relatively small value, $E=6$, which corresponds to the $v/c=0.99$ if the spacetime was asymptotically flat. 
From Fig. \ref{fig:sdsplot} (a), it is seen that for terms not involving $\Lambda$, i.e. $\Delta\phi_{\mathrm{S},n,0}~(n=1,2)$ plotted using the brown and red solid curves, the contribution for each $n$ decreases monotonically as $r_0$ increases, and the two terms 
is separated by about $M/r_0$. These are expected features similar to the perturbative deflection angle of Schwarzschild spacetime \cite{Jia:2020xbc}. For the terms containing $\Lambda$, one observes that for $\Delta\phi_{\mathrm{S},0,i}~(i=1,2)$, their sizes increase with the increase of $r_0$. For $\Delta\phi_{\mathrm{S},1,i}~(i=1,2)$ 
however, their sizes are almost independent of $r_0$ and then for $\Delta\phi_{\mathrm{S},2,i}~(i=1,2)$, their sizes decrease monotonically as $r_0$ increases. These features can be understood from the general order estimation Eq. \eqref{eq:dphiorderest}, where we knew that 
$\Delta\phi_{\mathrm{S},n,m}~(m\geq 1)$ to the leading order should be proportional to $r_0^{1-n}$. Among the terms dependent on $\Lambda$, clearly, $\Delta\phi_{\mathrm{S},0,1}$ is the largest. Comparing to the largest $\Lambda$ independent term $\Delta\phi_{\mathrm{S},1,0}$, it is seen that for the given choice of parameters $(M,r_{s,d},\Lambda)$, $\Delta\phi_{\mathrm{S},0,1}$ only becomes comparable when $r_0$ reaches about $8[^{\prime\prime}]\times r_d$. 

To study how good is the perturbative result \eqref{eq:dphiressch} in approximating the true deflection angle, we first defined a truncated $\Delta\phi_{\mathrm{S},\bar{n},\bar{m}}$ from \eqref{eq:dphiressch}
\be
\Delta\phi_{\mathrm{S},\bar{n},\bar{m}} =\sum_{j=s,d}
\sum_{n=0}^{\bar{n}}  \sum_{m=0}^{\bar{m}} \Delta\phi_{\mathrm{S},n,m}, \label{eq:dphitruncsch}
\ee
where $\bar{n}$ and $\bar{m}$ are the truncation orders of $n$ and $m$ respectively. And then in Fig. \ref{fig:sdsplot} (b), its relative difference $|\Delta\phi_{\mathrm{S},\bar{n},\bar{m}}-\Delta\phi_{\mathrm{S,num}}|/|\Delta\phi_{\mathrm{S,num}}|$ from the numerical deflection angle $\Delta\phi_{\mathrm{S,num}}$ obtained using numerical integration from the original definition \eqref{eq:dphidef} was plotted. 
It is seen that as both $\bar{n}$ and $\bar{m}$ increase, the percentage difference  between the truncated $\Delta\phi_{\bar{n}\bar{m}}$ and the $\Delta\phi_{\mathrm{S,num}}$ keeps decreasing. Note that because the $\Delta\phi_{\mathrm{S},n,m}~(m\geq 1)$ terms are negative, for some $\bar{n},\bar{m}$ and at some particular $r_0$, these truncated $\Delta\phi_{\mathrm{S},\bar{n},\bar{m}}$ might cross the numerical value 
$\Delta\phi_{\mathrm{S,num}}$ and therefore the difference at that point is zero. 

\subsection{RNdS Spacetime}
RNdS spacetime is described by the metric \eqref{eq:metric} with \eqref{eq:metricabform} and
\be 
f(r)=1-\frac{2M}{r}+\frac{Q^2}{r^2},~g(r)=1,~C(r)=r^2 \ee
As in the case of SdS spacetime, reading off the coefficients $f_i,~g_i$ and $c_i$ in Eq. \eqref{eq:fcexp} from this and substituting them into Eqs. \eqref{eq:pfirstfew} and \eqref{eq:alphaintres}, the change of the angular coordinate in the RNdS spacetime $\Delta\phi_{\mathrm{R}}$ to the order $(M/r_0)^2$ and $(r_0^2\Lambda)^1$ becomes
\begin{align}
\Delta\phi_{\mathrm{R}}=&\sum_{j=s,d} \Bigg[ \Delta\phi_{\mathrm{S},0,0}+\Delta\phi_{\mathrm{S},0,1}+\Delta\phi_{\mathrm{S},1,0}+\Delta\phi_{\mathrm{S},1,1} \nn \\
&+\frac{M^2}{r_0^2}  \lcb
-\frac12\left(\hat{Q}^2-3\right)  I_{2,-3} - \left(\hat{Q}^2-3\right)  I_{2,-2} \right.\nn\\
& -\frac12\lsb E^2\left(2  \hat{Q}^2-9 \right)+\kappa\left(3 - \hat{Q}^2\right)\rsb I_{2,-1}\nn\\
& -\frac{E^2}{ (E^2-\kappa )^2} \lsb E^2 \left(\hat{Q}^2-3 \right)+\kappa\left(5 -  \hat{Q}^2\right) \rsb I_{2,0}\nn\\
&\left.\left.-\frac{E^2}{2(E^2-\kappa )^2} \lsb E^2\left( \hat{Q}^2-3\right)+ \kappa\left( 4 -  \hat{Q}^2\right)\rsb I_{2,1}\rcb \rsb \nn\\
&+\mathcal{O}\lsb \lb \frac{M}{r_0}\rb^3,~(r_0^2\Lambda)^2\rsb
\label{eq:dphiresrn}
\end{align}
where $\hat{Q}\equiv Q/M$ and $\Delta\phi_{\mathrm{S},n,m}$ are given in Eq. \eqref{eq:dphisnm}. It is seen that the effect of the electric charge on the deflection angle emerges from order $1/r_0^2$ for the $\Lambda=0$ case. The large source/detector distance expansion of $\Delta\phi_\mathrm{R}$ is
\begin{align}
\Delta\phi^\prime_{\mathrm{R}}=& \sum_{j=s,d} \Bigg\{ \Delta\phi^\prime_{\mathrm{S},0,0}+\Delta\phi^\prime_{\mathrm{S},0,1}+\Delta\phi^\prime_{\mathrm{S},1,0}+\Delta\phi^\prime_{\mathrm{S},1,1} \nn \\
& +\frac{M^2}{r_0^2}
\frac{1}{ 8 (E^2-\kappa )^2} \lsb E^4\lb 15\pi-3\pi \hat{Q}^2-16\rb\right.\nn\\
&\left.+2E^2\kappa \lb 2\pi\hat{Q}^2-9\pi+4\rb +\lb 3-\hat{Q}^2\rb \pi\kappa \rsb \Bigg\} \nn\\
&+\mathcal{O}\lsb \lb \frac{M}{r_0}\rb^3,~(r_0^2\Lambda)^2,\lb \frac{r_0}{r_{s,d}}\rb^2\rsb
\label{eq:dphiresrninf}
\end{align}
where $\Delta\phi^\prime_{\mathrm{S},n,m}$ are given in Eqs. \eqref{eq:dphiresschinf}. We can easily verify that when setting $\hat{Q}=0$, the second order terms proportional to $(M/r_0)^2$ in Eqs. \eqref{eq:dphiresrn} and \eqref{eq:dphiresrninf} reduce to $\Delta\phi_{\mathrm{S},2,0}$ in Eq. \eqref{eq:dphis20} and $\Delta\phi^\prime_{\mathrm{S},2,0}$ in Eq. \eqref{eq:dphiresschinf20}. On the other hand, taking $\Lambda=0$, Eqs. \eqref{eq:dphiresrn} and \eqref{eq:dphiresrninf} reduce to the corresponding results in RN spacetime for both null and timelike rays \cite{Duan:2020tsq,Huang:2020trl}. 

From Eq. \eqref{eq:dphiresrninf} it is clear that the spacetime charge $Q$ does not appear in the first order terms of $(r_0^2\Lambda)$. Therefore to this order, the effect of $\Lambda$ on $\Delta\phi_{\mathrm{R}}$ is the same as its effect in SdS spacetime. As we saw in Subsec. \ref{subsec:sds} for the SdS case, the first order of $(r_0^2\Lambda)$ is already very small and therefore we will not pursue to study the second order terms proportional to $(r_0^2\Lambda)^2$ in the RNdS spacetime.

\subsection{Dilaton-dS Spacetime}

The metric of the dilaton-dS spacetime is given by Eq.  \eqref{eq:metric} with \eqref{eq:metricabform} and \cite{Gao:2004tu}
\be 
f(r)=1-\frac{2M}{r},~g=1-\frac{2D}{r},~C(r)=r^2\lb 1-\frac{2D}{r}\rb 
\label{eq:ddsmetric}
\ee
where $D$ is the dilaton charge. When $D=0$, this reduces to the SdS spacetime metrics. Reading off the coefficients $f_i,~g_i$ and $c_i$ in Eq. \eqref{eq:fcexp} from this and substituting into Eqs. \eqref{eq:pfirstfew} and  \eqref{eq:alphaintres}, we have $\Delta\phi_{\mathrm{D}}$ in this spacetime to the leading orders
\begin{align} 
\Delta\phi_{\mathrm{D}}=&
\sum_{j=s,d} \lsb \Delta\phi_{\mathrm{D},0,0} +
\Delta\phi_{\mathrm{D},0,1} +\Delta\phi_{\mathrm{D},1,0} +\Delta\phi_{\mathrm{D},1,1}\right.\nn\\
&\left.+\Delta\phi_{\mathrm{D},2,0} \rsb+\mathcal{O}\lsb \lb \frac{M}{r_0}\rb^3,~(r_0^2\Lambda)^2\rsb
\label{eq:dphiresdilaton}
\end{align}
where 
\begin{subequations}
\label{eq:dphidnm}
\begin{align}
\Delta\phi_{\mathrm{D},0,0}=& \Delta\phi_{\mathrm{S},0,0},\\
\Delta\phi_{\mathrm{D},0,1}=&\Delta\phi_{\mathrm{S},0,1}, \label{eq:dphid01}\\
\Delta\phi_{\mathrm{D},1,0}=& \frac{M}{r_0} \lsb  (1+\hat{D})\lb I_{1,-2}+ I_{1,-1}\rb\right.\nn\\
&\left.+\lb \frac{ E^2}{E^2-\kappa}-\hat{D}\rb I_{1,0} \rsb, \\
\Delta\phi_{\mathrm{D},1,1}=&\frac{M}{r_0}(r_0^2 \Lambda)\frac{ \kappa }{6 \left(E^2-\kappa \right)} \times \lsb \lb \hat{D}-\frac{ 5 E^2-3  }{ E^2-\kappa } \rb I_{1,0} \right. \nn \\ 
&\left. +\lb \hat{D}-\frac{ E^2-3  }{ E^2-\kappa } \rb I_{1,1} +\lb \hat{D}-\frac{E^2 }{E^2-\kappa} \rb I_{1,2} \rsb,\label{eq:dphid11}\\
\Delta\phi_{\mathrm{D},2,0}=&\frac{M^2}{r_0^2}
 \lcb \frac{1}{2} \left(3 \hat{D}^2+2 \hat{D}+3\right)  I_{2,-3}\right.\nn\\
&+(3 \hat{D}^2+2 \hat{D}+3)  I_{2,-2}  \nn \\
&+\lsb \frac{\hat{D}^2}{2}-\frac{\hat{D} E^2}{E^2-\kappa }+\frac{3 \left(3 E^2-\kappa \right)}{2 \left(E^2-\kappa \right)} \rsb I_{2,-1}\nn\\
&+ \lsb -3 \hat{D}^2+\frac{\hat{D} \kappa }{E^2-\kappa }+\frac{E^2 \left(3 E^2-5 \kappa \right)}{\left(E^2-\kappa \right)^2} \rsb I_{2,0} \nn \\
&\left.+\lsb -\frac{\hat{D}^2}{2}-\frac{\hat{D} E^2}{E^2-\kappa }+\frac{E^2 \left(3 E^2-4 \kappa \right)}{2 \left(E^2-\kappa \right)^2} \rsb I_{2,1} \rcb
\end{align}
\end{subequations}
where $\Delta\phi_{\mathrm{S},0,0}$ and $\Delta\phi_{\mathrm{S},0,1}$ are given in Eqs. \eqref{eq:dphis00} and \eqref{eq:dphis01} and $\hat{D}\equiv D/M$. The finite distance expansion of these contributions are
\begin{subequations}
\label{eq:dphidexp}
\begin{align}
\Delta\phi^\prime_{\mathrm{D},0,0}=&  \Delta\phi^\prime_{\mathrm{S},0,0},\\
\Delta\phi^\prime_{\mathrm{D},0,1}=&\Delta\phi^\prime_{\mathrm{S},0,1},\\
\Delta\phi^\prime_{\mathrm{D},1,0}=&\frac{M}{r_0}  \lsb  \left(\hat{D}-\frac{E^2}{E^2-\kappa }\right)\frac{r_0}{r_j}+\frac{2 E^2-\kappa}{E^2-\kappa } \rsb ,  \\
\Delta\phi^\prime_{\mathrm{D},1,1}=&\frac{M}{r_0} (r_0^2 \Lambda) \frac{\kappa }{6  \left(E^2-\kappa \right)} \lcb \lsb \hat{D} -\frac{E^2}{\left(E^2-\kappa \right)}\rsb \frac{r_j}{r_0} \right.  \nn \\ 
&\left.+\hat{D} -\frac{ 5E^2-3 \ln \lb \frac{2r_j}{r_0} \rb }{E^2-\kappa} \rcb,\\
\Delta\phi^\prime_{\mathrm{D},2,0}=&\frac{M^2}{r_0^2} \lsb -\frac{1}{8} \pi  \hat{D}^2-\frac{\hat{D} \left((3 \pi  -8) E^2+(4-\pi)  \kappa \right)}{4 \left(E^2-\kappa \right)} \right. \nn \\ &\left. +\frac{ (15 \pi -16) E^4+(8-18 \pi)  E^2 \kappa +3 \pi  \kappa }{8 \left(E^2-\kappa \right)^2} \rsb
\end{align}
\end{subequations}
where $\Delta\phi^\prime_{\mathrm{S},0,0}$ and $\Delta\phi^\prime_{\mathrm{S},0,1}$ are given in Eqs. \eqref{eq:dphiresschinf00} and \eqref{eq:dphiresschinf01}. When the dilaton charge is zero, Eqs. \eqref{eq:dphidnm} and \eqref{eq:dphidexp} reduce to the corresponding SdS results \eqref{eq:dphisnm} and \eqref{eq:dphiresschinf} respectively.

Unlike the charge parameter $Q$ in the RNdS spacetime, the dilaton charge $D$ in $\Delta\phi_{\mathrm{D}}$ appears from the first order of $M/r_0$, i.e., in the $\Delta\phi_{\mathrm{D},1,m}$ terms in Eq. \eqref{eq:dphidnm} and $\Delta\phi^\prime_{\mathrm{D},1,m}$ terms in Eq. \eqref{eq:dphidexp}. For the effect of $\Lambda$ in which we are more interested however, because the metric functions \eqref{eq:ddsmetric} of the dilaton-dS spacetime also satisfy the condition \eqref{eq:metriccform}, it is seen from Eqs. \eqref{eq:dphid01} and \eqref{eq:dphid11} that when $\kappa=0$, the $(r_0^2\Lambda)$ terms will also vanish completely, as predicted in Sec. \ref{subsec:nullindep}. Therefore, the effect of $\Lambda$ on $\Delta\phi_{\mathrm{D}}$ of null rays with fixed $r_0$ will be absent too.

\subsection{Brane-World BH Spacetime}

The line element of the Brane-World BH is given by Eq. \eqref{eq:metric} with \eqref{eq:metricabform} and \cite{Jalalzadeh:2012mw}
\be 
f(r)=1-\frac{2M}{r},~g=1+\frac{2\hat{\beta}}{r}+\frac{\hat{\beta}^2}{r^2} ,~C(r)=r^2, \label{eq:bwmetric}
\ee
with $\Lambda=\alpha^2,~\hat{\beta}\equiv \beta/\alpha$ and $\alpha,~\beta$ are some parameters that have not been well constrained. When $\hat{\beta}=0$, clearly this reduces to the SdS spacetime.
Reading off the coefficients and substituting them into Eqs. \eqref{eq:pfirstfew} and  \eqref{eq:alphaintres},  the change of the angular coordinate in this spacetime becomes
\begin{align} 
\Delta\phi_{\mathrm{B}}=&
\sum_{j=s,d} \lsb \Delta\phi_{\mathrm{B},0,0} +
\Delta\phi_{\mathrm{B},0,1} +\Delta\phi_{\mathrm{B},1,0} +\Delta\phi_{\mathrm{B},1,1}\right.\nn\\
&\left.+\Delta\phi_{\mathrm{B},2,0} \rsb+\mathcal{O}\lsb \lb \frac{M}{r_0}\rb^3,~(r_0^2\Lambda)^2\rsb.
\label{eq:dphiresbw}
\end{align}
where the contributions from each order are
\begin{subequations}
\label{eq:dphibwnm}
\begin{align}
\Delta\phi_{\mathrm{B},0,0}=& \Delta\phi_{\mathrm{S},0,0},\\
\Delta\phi_{\mathrm{B},0,1}=&\Delta\phi_{\mathrm{S},0,1}, \\
\Delta\phi_{\mathrm{B},1,0}=& \Delta\phi_{\mathrm{S},1,0}, \\
\Delta\phi_{\mathrm{B},1,1}=&\frac{M}{r_0}(r_0^2 \Lambda)\times \lcb \lsb -\frac{\kappa  \left(5 E^2-3  \right)}{6 \left(E^2-\kappa \right)^2}+\frac{\hat{\beta }}{3 M} \rsb I_{1,0} \right. \nn\\ 
&  -\lsb \frac{\kappa  \left(E^2-3  \right)}{6 \left(E^2-\kappa \right)^2}+\frac{\kappa \hat{\beta }  }{3 M \left(E^2-\kappa \right)} \rsb I_{1,1}\nn\\
&\left.-\frac{\kappa E^2}{6 \left(E^2-\kappa \right)^2} I_{1,2} \rcb,\\
\Delta\phi_{\mathrm{B},2,0}=&\Delta\phi_{\mathrm{S},2,0}
\end{align}
\end{subequations}
where $\Delta\phi_{\mathrm{S},n,m}$ are given in Eq. \eqref{eq:dphisnm}. Note for the terms in Eq. \eqref{eq:dphiresbw} that do not involve $(r_0^2\Lambda)$, i.e., terms except $\Delta\phi_{\mathrm{B},n,1}$, their forms are exactly as in the SdS case because in the case that $\Lambda$ is not present, substituting Eq. \eqref{eq:bwmetric} into Eq. \eqref{eq:metricabform} leads exactly to the SdS spacetime.

Again, to see the finite distance effect, using the expansions \eqref{eq:i1nexp} for the first few $I_{n,k}$'s,  the $\Delta\phi_{\mathrm{B},n,m}$ become to the leading orders of $r_0/r_i~(i=s,d)$
\begin{subequations}
\label{eq:dphibwexp}
\begin{align}
\Delta\phi^\prime_{\mathrm{B},0,0}=&  \Delta\phi^\prime_{\mathrm{S},0,0},\\
\Delta\phi^\prime_{\mathrm{B},0,1}=&\Delta\phi^\prime_{\mathrm{S},0,1},\\
\Delta\phi^\prime_{\mathrm{B},1,0}=&\Delta\phi^\prime_{\mathrm{S},1,0},  \\
\Delta\phi^\prime_{\mathrm{B},1,1}=&\frac{M}{r_0} (r_0^2 \Lambda)\times \lcb -\frac{\kappa E^2}{6(E^2-\kappa)^2}\frac{r_j}{r_0} +\frac{E^2 \frac{\hat{\beta}}{M}}{3(E^2-\kappa)}\right. \nn\\ 
&\left. +\frac{\kappa\lsb \lb(2-2E^2)\frac{\hat{\beta}}{M} +3 \rb \ln(\frac{2r_j}{r_0})-5 E^2\rsb }{6(E^2-\kappa)^2} \rcb,\\
\Delta\phi^\prime_{\mathrm{B},2,0}=&\Delta\phi^\prime_{\mathrm{S},2,0},
\end{align}
\end{subequations}
where again  $\Delta\phi^\prime_{\mathrm{S},n,m}$ are given in Eq. \eqref{eq:dphiresschinf}.
A key difference of Eq. \eqref{eq:dphibwexp} from the previous SdS, RNdS and dilaton-dS spacetime results is that even for the null signal, the effect of the cosmological constant $\Lambda$ will still be present. That is, for light, we have
\be \Delta\phi^\prime_{B,1,1}(\kappa=0)
=\frac{M}{r_0}(r_0^2\Lambda)\frac{\hat{\beta}}{3M}=\frac{\hat{\beta}r_0\Lambda}{3}. 
\ee
The reason is simply that the condition \eqref{eq:metriccform} is broken by the metric \eqref{eq:bwmetric} of the Brane-World spacetime. 

\begin{figure}[htp!]
    \centering
\includegraphics[width=0.45\textwidth]{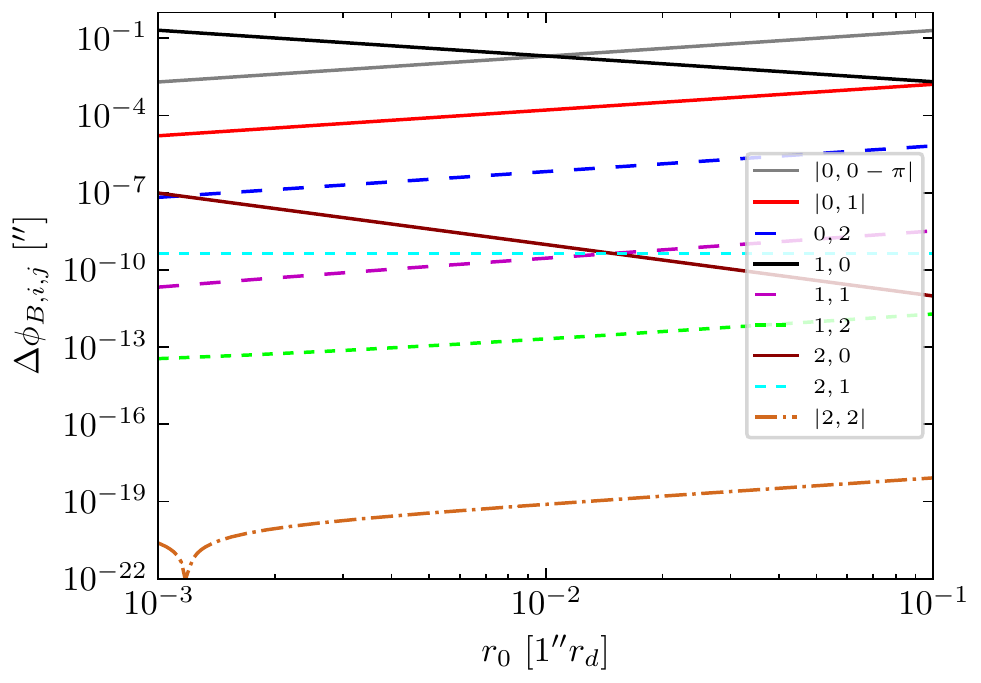}\\
    (a)\\
    \includegraphics[width=0.45\textwidth]{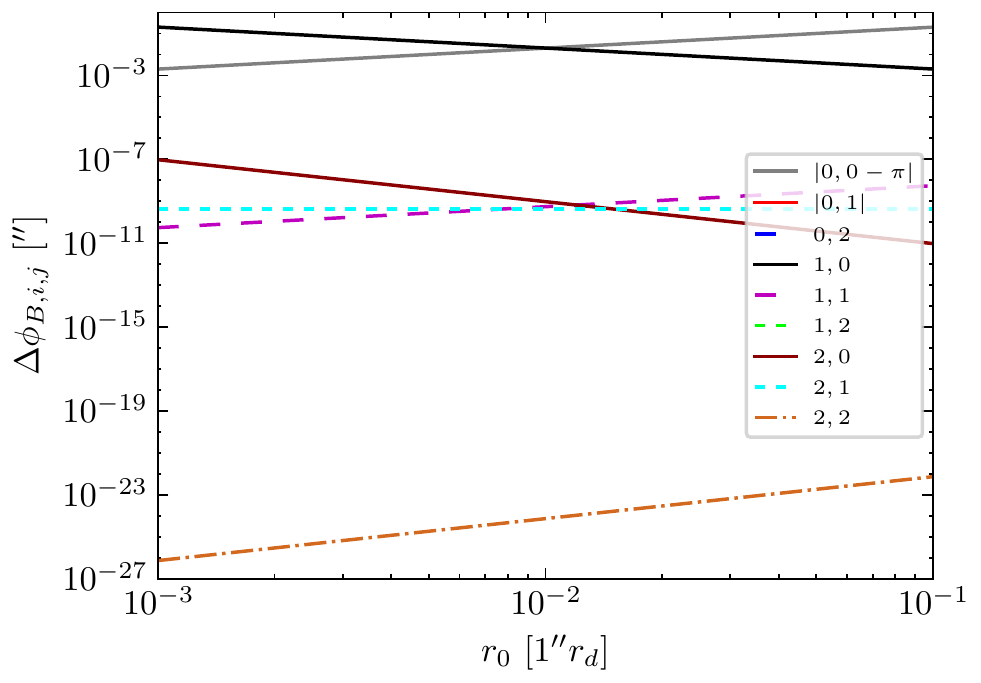}\\
    (b)
    \caption{$\Delta\phi_{\mathrm{B}}$ using Eq. \eqref{eq:dphibwexp}. (a) $\kappa=1,E=6$; (b) $\kappa=0$. Other parameters are the same as in Fig. \ref{fig:sdsplot}.  \label{fig:bwplot}}
\end{figure}

In Fig. \ref{fig:bwplot} (a) and (b), we plot the contribution from each order of Eq. \eqref{eq:dphibwexp} for $(\kappa=1, E=6)$ and $\kappa=0$ respectively. We chose the parameter $\hat{\beta}=0.01^{\prime\prime}r_d$. Other parameters are the same as in Fig.  \ref{fig:sdsplot}. Comparing plots (a) and (b), we observe that unlike the case of $\Delta\phi_{\mathrm{S}}$, some of the $\Delta\phi_{\mathrm{B},n,m}~(m\geq 1)$  terms proportional to $\Lambda^m$
also survived after setting $\kappa=0$, e.g. the $\Delta\phi_{\mathrm{B},1,1},~\Delta\phi_{\mathrm{B},2,1}$ and $\Delta\phi_{\mathrm{B},2,2}$ terms.

\section{The Gravitational Lensing in asymptotically dS spacetimes\label{sec:glcases}}

\subsection{The GL apparent angles}

In order to study the GL effect in asymptotically dS spacetimes,  we should first build a proper lensing equation (LE), from which the apparent angles of the lensed images can be solved for any given angular position of the lens, i.e., $\beta_L$ or $\beta$ in Fig. \ref{fig:sdsf1}. Traditionally, this LE was built using deflection angle for source and observer located at spacial infinity, together with some approximate geometric relation.
However in the case that spatial infinity can not be reached as in asymptotically dS spacetimes, this is not possible. In contrast, if $\Delta\phi$ with finite distance effect is known, we can actually build an exact LE without having to set the source/detector to some approximate locations. 
This exact LE is just the relation between the change of angular coordinate $\Delta \phi$ given in Eq. \eqref{eq:alphaintres} and the source angular location $\beta_L$ (see Fig. \ref{fig:sdsf1})
\be 
\pi\pm \beta_L=\Delta\phi (r_0)  ,\label{eq:glinr0}\ee 
where the $+$ and $-$ signs correspond respectively to the trajectories rotating anti-clockwise and clockwise. 
From this equation, the closest distance $r_0$ can be solved, after which using Eqs. \eqref{eq:ler0rel} and \eqref{eq:aasol} the apparent angle $\theta$ can be obtained.

Since $\Delta\phi$ in \eqref{eq:alphaintres}  after the large $r_{s,d}/r_0$ expansion of $I_{n,k}$ is a rational function of $r_0$, the analytical solution to Eq. \eqref{eq:glinr0} will be difficult if we keep too high or too many orders of $r_0$ in $\Delta\phi$. Therefore we will truncate $\Delta\phi$ in Eq. \eqref{eq:glinr0} to the highest solvable order. We will analyze the GL in two limiting cases. Case (1) is when
$M/r_0$ is as large as possible while $r_{s,d}^2\Lambda$ is as small as possible, and case (2) is when $M/r_0$ is as small as possible while $r_{s,d}^2\Lambda$ is as large as possible. 

In case (1), such as in the GL by the SgrA* SMBH, we would have a typical $\mathcal{O}(M/r_0)=10^{-6}\sim10^{-5}$, and if $\Lambda=\Lambda_c$ is used, $r_{s,d}^2\Lambda_c=6.7\times 10^{-12}$. Then the $n=0,1$ and $m=0$ terms in $\Delta\phi$ are much larger than the $n\geq 2,~m=0$ and $n\geq 0,~m\geq1$ terms and therefore keeping only the former terms, the LE \eqref{eq:glinr0} becomes 
\begin{align}
\pi\pm \beta_L=&\sum_{j=s,d}
\lsb p_{0,0,-1} I_{0,-1}\lb \frac{r_j}{r_0}\rb \right.\nn\\
&~~~~~~~\left.
+\frac{1 }{r_0 } \sum_{k=-2}^{0} p_{1,0,k} I_{1,k}\lb \frac{r_j}{r_0}\rb 
\rsb
. \label{eq:gleqcase1}
\end{align}
To solve $r_0$, we will directly use the large $r_{s,d}/r_0$ expansions of $I_{0,-1}$ and $I_{1,k}~(k=-2,-1,0)$ given in Eq. \eqref{eq:i1nexp}. Then the above equation becomes a quadratic equation of $r_0$
\begin{align}
    \pi\pm \beta_L=&
\sum_{j=s,d}
\lcb \frac\pi2 -\frac{r_0}{r_j} 
+\frac{1 }{r_0 } \lsb p_{1,0,-2} \lb 2-\frac\pi2\rb\right.\right.\nn\\
&~~~~~~~\left.\left.+p_{1,0,-1}\lb \frac\pi2-1\rb
+p_{1,0,0}\rsb 
\rcb\nn\\
=&
\sum_{j=s,d}
\lsb \frac\pi2 -\frac{r_0}{r_j} 
-\frac{1 }{r_0 } \frac{(2E^2-\kappa)f_1}{2(E^2-\kappa)}
\rsb
\end{align}
where in the second step $p_{0,0,-1}$ and $p_{1,0,k}$ in Eq. \eqref{eq:pfirstfew} were substituted.
Solving this, we find two allowed $r_0$: $r_{0+}$ and $r_{0-}$ for signals from the counterclockwise and clockwise directions respectively 
\begin{align}
r_{0\pm}=& \frac{ 1}{2 ( r_d + r_s )}\lsb  \mp  r_d r_s \beta_L \right. \nn\\
&\left. +\sqrt{ r_d^2 r_s^2 \beta_L^2 -\frac{4 r_d r_s ( r_d + r_s ) (2E^2-\kappa)f_1}{E^2-\kappa }}\rsb.\label{eq:r0solcase1}
\end{align}

Case (2) can be realized by the GL of more distant objects so that $r_{s,d}$ and consequently $r_{s,d}^2 \Lambda$ can be larger. In this case,  the $n=0,~m=0,1$ terms of $\Delta\phi$ could be more important than the $n=1,~m=0$ term which is certainly much larger than the $n\geq 2$ terms. Therefore the LE \eqref{eq:glinr0} becomes 
\begin{align}
\pi\pm \beta_L=&\sum_{j=s,d}
\lsb p_{0,0,-1} I_{0,-1}\lb \frac{r_j}{r_0}\rb
+\frac{1 }{r_0 } \sum_{k=-2}^{0} p_{1,0,k} I_{1,k}\lb \frac{r_j}{r_0}\rb \right. 
\nn\\
&\left.
+(r_0^2\Lambda) \sum_{k=-1}^{1} p_{0,1,k} I_{0,k}\lb \frac{r_j}{r_0}\rb 
\rsb
, \label{eq:gleqcase2}
\end{align}
Here we kept the term proportional to $1/r_0$ because this does not spoil the solvability of the equation. 
Using the large $r_{s,d}/r_0$ expansion of $I_{0,k}~(k=-2,-1,0)$ and $I_{1,k}~(k=-1,0,1)$ in Eq. \eqref{eq:i1nexp}, this also becomes a quadratic equation of $r_0$,
\begin{align}
\pi\pm \beta_L=&\sum_{j=s,d}
\lcb \frac\pi2 -\frac{r_0}{r_j} 
+\frac{1 }{r_0 } \lsb p_{1,0,-2} \lb 2-\frac\pi2\rb \right.\right.\nn\\
&\left.\left.+p_{1,0,-1}\lb \frac\pi2-1\rb
+p_{1,0,0}\rsb 
+r_0r_j\Lambda p_{0,1,1}\rcb\nn\\
=&\sum_{j=s,d}
\lsb \frac\pi2 -\frac{r_0}{r_j} 
-\frac{1 }{r_0 } \frac{(2E^2-\kappa)f_1}{2(E^2-\kappa)}
-\frac{\kappa r_0r_j\Lambda}{6(E^2-\kappa)} \rsb
\end{align}
where in the second step the $p_{n,m,k}$ were substituted by Eq. \eqref{eq:pfirstfew}. From this, again $r_0$ can be readily solved as
\begin{align}
&r_{0\pm}= \frac{ 1}{2 ( r_d + r_s )\lsb 1+\kappa r_sr_d\Lambda/[6(E^2-\kappa)]\rsb }\Bigg[  \mp  r_d r_s \beta_L + \nn\\
&\left. \sqrt{ r_d^2 r_s^2 \beta_L^2 -\frac{4 r_d r_s ( r_d + r_s ) (2E^2-\kappa)f_1}{E^2-\kappa }\lb 1+\frac{\kappa r_sr_d\Lambda}{6(E^2-\kappa)}\rb }\rsb.\label{eq:r0solcase2}
\end{align}
The $\beta_L$ in this and the former result \eqref{eq:r0solcase1} can be replaced by the more commonly used $\beta$ in Fig. \ref{fig:sdsf1}, by using the following geometrical relation between them
\be 
r_s\sin \beta_L=(r_d+r_s\cos\beta_L)\tan\beta .
\ee 
This equation allows us to rewrite any formula involving $\beta_L$ into a formula using $\beta$. Solving this to the first order and substituting into Eq. \eqref{eq:r0solcase2}, $r_{0\pm}$ become 
\begin{align}
&r_{0\pm}= \frac{ r_d}{2\lsb 1+\kappa r_sr_d\Lambda/[6(E^2-\kappa)]\rsb }\Bigg[  \mp  \beta + \nn\\
&\left. \sqrt{ \beta^2 -\frac{4 r_s  (2E^2-\kappa)f_1}{r_d ( r_d + r_s )(E^2-\kappa )}\lb 1+\frac{\kappa r_sr_d\Lambda}{6(E^2-\kappa)}\rb }\rsb.\label{eq:r0solcase2inbeta}
\end{align}

Clearly, Eq. \eqref{eq:r0solcase2inbeta}
is a more general result than  Eq. \eqref{eq:r0solcase1}, which is the asymptotically flat spacetime result: when $\Lambda\to 0$, Eq. \eqref{eq:r0solcase2inbeta} reduces to Eq. \eqref{eq:r0solcase1}. In order for the effect of \mclambda on $r_{0\pm}$ to be apparent, from Eq. \eqref{eq:r0solcase2inbeta} we see that we should have for $\kappa=1$ the following condition
\be r_sr_d\Lambda\gtrsim 6(E^2-1).\label{eq:lambdaonr0cond}
\ee
Noting the value of $\Lambda_c$, this means only for source and lens that are of cosmic distance can the effect of \mclambda be important to $r_{0\pm}$. When this is the case, expanding \eqref{eq:r0solcase2} in small $\Lambda$, one arrives at
\begin{align}
&r_{0\pm}(\Lambda\to0)=r_{0\pm}(\Lambda=0)+\Lambda\times\lsb \frac{r_dr_s\beta_L}{2(r_d+r_s)}
\right.\nn\\
&\left.-\frac{r_d^2 r_s^2 \beta_L^2 -\frac{2 r_d r_s ( r_d + r_s ) (2E^2-\kappa)f_1}{E^2-\kappa }}{2(r_d+r_s)\sqrt{ r_d^2 r_s^2 \beta_L^2 -\frac{4 r_d r_s ( r_d + r_s ) (2E^2-\kappa)f_1}{E^2-\kappa }}}\rsb, \label{eq:r0smalllexp}
\end{align}
where $r_{0\pm}(\Lambda=0)$ is just the $r_{0\pm}$ in Eq. \eqref{eq:r0solcase1}. It is not too difficult by using typical parameter values to numerically verify that the square bracket part is actually negative, and therefore a positive \mclambda will decrease both $r_{0+}$ and $r_{0-}$. 

Having obtained $r_{0\pm}$, substituting it into Eq. \eqref{eq:aasol}, we can get the apparent angles of the two images 
\be
\theta_\pm=\arcsin\lb \sqrt{\frac{E^2-\kappa A(r_{0\pm})}{E^2-\kappa A(r_d)}}\sqrt{\frac{A(r_d)C(r_{0\pm})}{A(r_{0\pm})C(r_d)}}\rb.
\label{eq:aasolwithr0pm}\ee
Since $r_d\gg r_0\gg |f_1|=2M$, and if $\mathcal{O}(f_n,g_n,c_n)\leq M^n$, then these apparent angles can be further expanded in terms of the ratios among $\{r_d,~r_0,~M,~\Lambda\}$. Substituting Eqs. \eqref{eq:metricabform}, \eqref{eq:fcexp} into \eqref{eq:aasolwithr0pm}, carrying out the expansion in terms of $\Lambda, ~r_{0\pm}/r_d, ~M/r_{0\pm}$ and keeping only to the first order for each of these, the apparent angles become
\begin{align} 
\theta_\pm=&\frac{r_{0\pm}}{r_d}-\frac{\Lambda r_{0\pm} r_d E^2}{6(E^2-\kappa)}+\mathcal{O}\lsb \lb \frac{r_0}{r_d}\rb^2,\Lambda r_dM, \Lambda^2\rsb
\end{align}
where $r_{0\pm}$ is in Eq. \eqref{eq:r0solcase2}.
Further carrying out the small \mclambda expansion in $r_{0\pm}$ using \eqref{eq:r0smalllexp}, the effect of \mclambda on $\theta_\pm$ becomes more apparent
\begin{align}
\theta_\pm=&\frac{r_{0\pm}(\Lambda=0)}{r_d}+\lsb \frac{C_\Lambda}{r_d}-\frac{r_{0\pm}(\Lambda =0) E^2}{6(E^2-\kappa)}\rsb \Lambda\nn\\ &+\mathcal{O}\lsb \lb \frac{r_0}{r_d}\rb^2,\Lambda r_dM, \Lambda^2\rsb\label{eq:thetapmlexp}
\end{align}
where $C_\Lambda$ is the coefficient of $\Lambda$ in the second term of Eq. \eqref{eq:r0smalllexp}. Since both terms in the square bracket are negative, it is clear then $\theta_\pm$ will also decrease as $\Lambda $ increases. We point out that if $\Lambda=0$ as in asymptotically flat spacetimes,  the $r_{0\pm}(\Lambda=0)/r_d$ term can be re-expressed using Eq. \eqref{eq:r0solcase1}
in terms of the angle $\beta$, and $\theta_\pm(\Lambda=0)$ become
\be 
\theta_{\pm}(\Lambda=0)=\frac{r_{0\pm}(\Lambda=0)}{r_d}=\frac{1}{2}\lb \sqrt{\beta^2
 +4\theta_{\mathrm{E}}^2}\mp \beta\rb ,\label{eq:thetapml0}
\ee 
where the Einstein angular radius $\theta_{\mathrm{E}}(\Lambda=0)$ is
\be 
\theta_{\mathrm{E}}(\Lambda=0)=\sqrt{
-\frac{4 r_s  (2E^2-\kappa)f_1}{r_d( r_d + r_s )(E^2-\kappa) }}. \ee
Eq. \eqref{eq:thetapml0} agrees perfectly with the apparent angles of conventional point mass lensing \cite{bk:keeton}. 

\begin{figure}[htp!]
    \centering
    \includegraphics[width=0.4\textwidth]{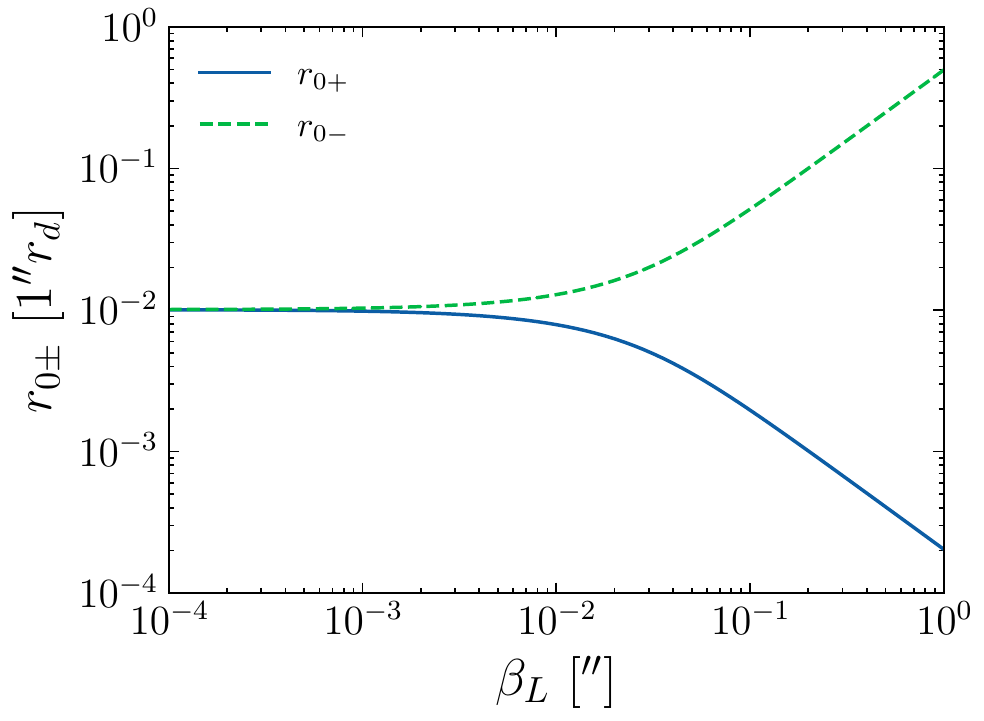}\\
    (a)\\
    \includegraphics[width=0.4\textwidth]{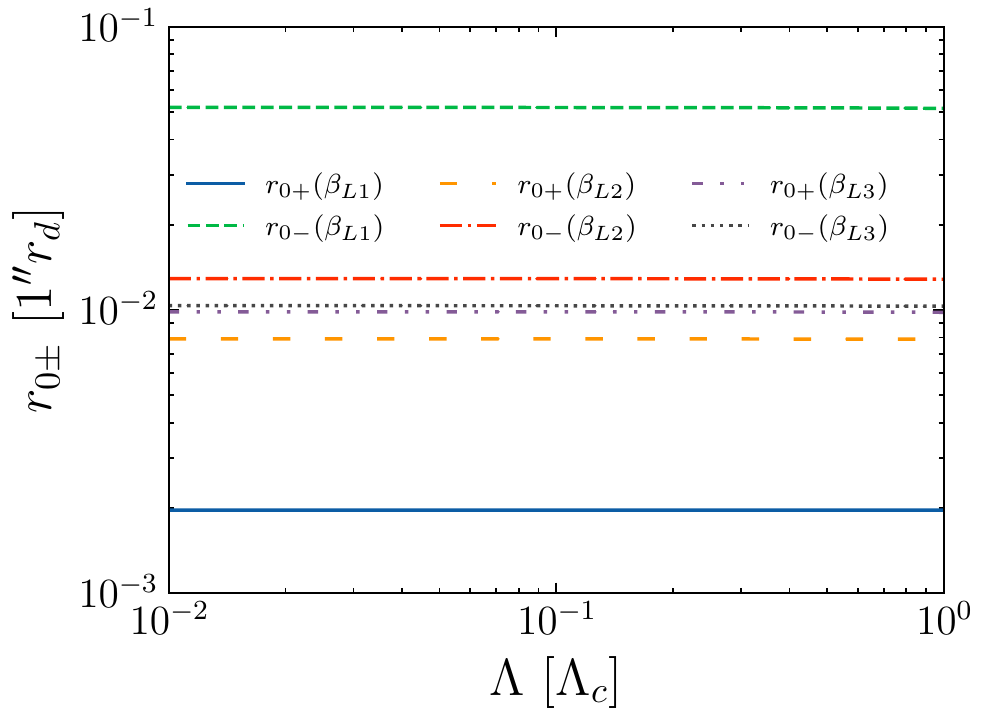}\\
    (b)\\
    \includegraphics[width=0.4\textwidth]{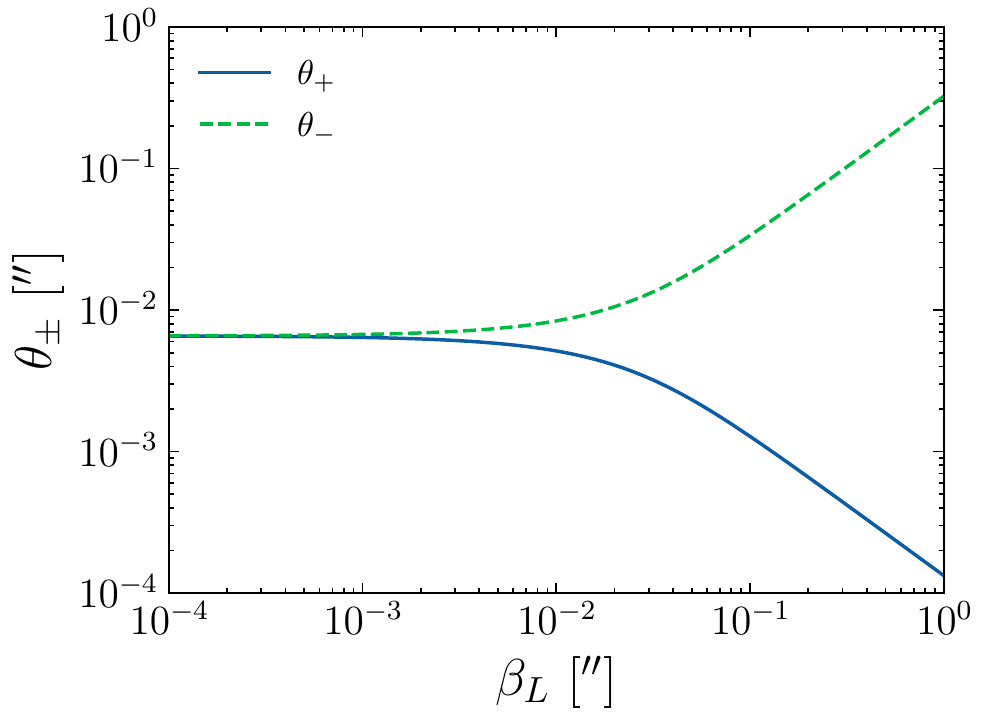}\\
    (c)\\
    \includegraphics[width=0.4\textwidth]{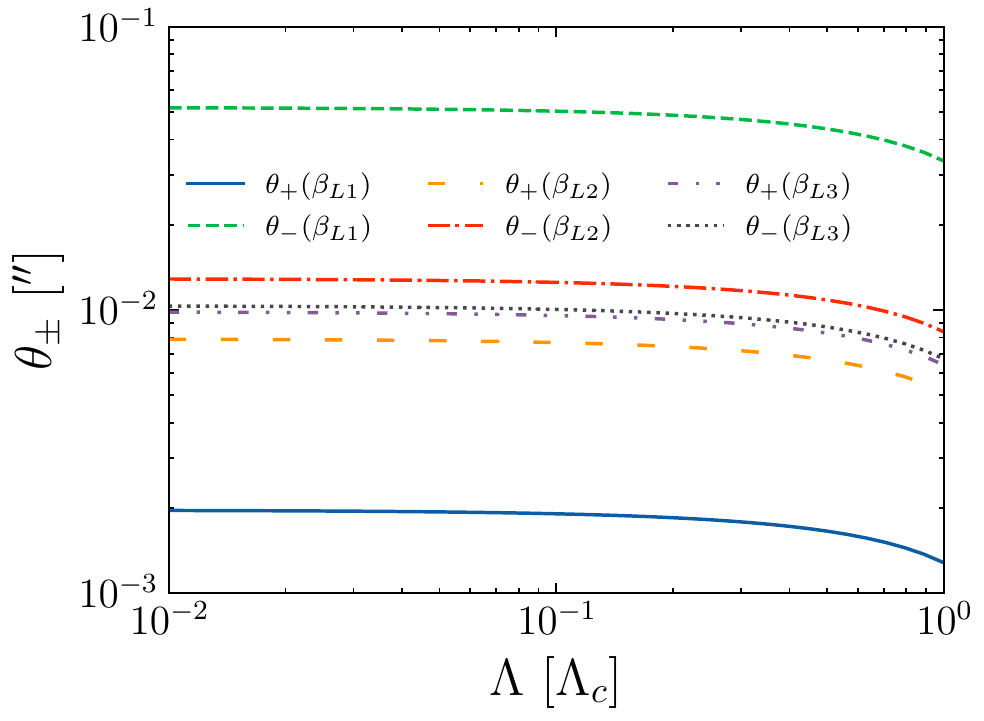}\\
    (d)
    \caption{The minimal radii $r_{0\pm}$ and apparent angles $\theta_{\pm}$ for two trajectories from two sides of the lens, as functions of $\beta_L$ ((a) and (c)) and $\Lambda$ ((b) and (d)). The $\Lambda$ in (a) and (c) are set to $\Lambda_c$ and the chosen $\beta_{L1}$ to $\beta_{L3}$ in (b) and (d) are respectively $10^{-3},~10^{-2}$ and $10^{-1}$  [$^{\prime\prime}$]. }
    \label{fig:r0thetaplot}
\end{figure}

In Fig. \ref{fig:r0thetaplot}, we plot $r_{0\pm}$ in  \eqref{eq:r0solcase2} and the two associated apparent angles $\theta_\pm$ in \eqref{eq:aasolwithr0pm} as functions of $\beta_L$ and $\Lambda$. The coefficient $f_1$ is set to $-2M$ and other parameters are the same as in Fig. \ref{fig:sdsplot}. We see from Fig. \ref{fig:r0thetaplot} (a) that with this parameter setting, the value of $r_{0\pm}(\beta_L\to0)/r_d$, which is roughly the Einstein radius (in asymptotic flat spacetime), is 0.010 [$^{\prime\prime}$] and the minimal radius $r_{0+}$ (or $r_{0-}$) decreases (or increases) as $\beta_L$ increases, as expected from Fig. \ref{fig:sdsf1}. 
From Fig. \ref{fig:r0thetaplot} (c) we see however, the exact value of $\theta_\mathrm{E}$ (defined as the value of $\theta_{\pm}$ at $\beta_L=0$) is at 0.0066 [$^{\prime\prime}$], slightly different from the value estimated from Fig. \ref{fig:r0thetaplot} (a). The reason is that the estimation $\theta_\mathrm{E}\approx r_{0\pm}(\beta_L\to 0)/r_d$ only works well when $\Lambda=0$, as indicated by Eqs. \eqref{eq:thetapmlexp} and \eqref{eq:thetapml0}, and the extra term proportional to \mclambda in Eq. \eqref{eq:thetapmlexp} decreases this value. 

On the other hand,  from Fig. \ref{fig:r0thetaplot} (b) we observe that for all three chosen $\beta_L$ ($10^{-3},~10^{-2}$ and $10^{-1}$  [$^{\prime\prime}$]), $r_{0\pm}$ are (almost) independent of $\Lambda$. As pointed out earlier, this is because for the chosen $r_s,~r_d,~\Lambda$ and $E$, the condition  \eqref{eq:lambdaonr0cond} is still far from being reached, which consequently suppressed the effect of $\Lambda$. As for the dependence of $\theta_\pm$ on $\Lambda$, we see from Fig. \ref{fig:r0thetaplot} (d) that unlike $r_{0\pm}$, the apparent angles from both sides of the lens decrease very apparently as $\Lambda$ increases, as predicted by Eq. \eqref{eq:thetapmlexp}. 

\subsection{Time delay between GL images\label{subsec:td}}

With $r_{0\pm}$ known in Eq. \eqref{eq:r0solcase2inbeta}, we can now utilize the total travel time \eqref{eq:totaltfirstfew} to compute the time delay between GL images. If the images are formed by the rays of the same energy $E$ but from different sides of the lens, then such time delay is
\begin{align}
\Delta^2 t_\pm \equiv &\Delta t(r_{0+}(E),E) -\Delta t(r_{0-}(E),E),\label{eq:tdpmdef}
\end{align}
where both the minimal radii $r_{0\pm}$ and total travel time explicitly depend on the signal energy. 
On the other hand, if we consider the time delay between signals from the same side of the lens but having different energies, such as the time delay between light (or GW) and neutrinos, or neutrinos of different mass eigenstates, then such time delay is
\begin{align}
\Delta^2 t_{12} \equiv &\Delta t(r_{0\xi}(E_1),E_1) -\Delta t(r_{0\xi}(E_2),E_2), \label{eq:td12def}
\end{align}
where $\xi=\pm$ represents the signals from the ``$+$'' or ``$-$'' side of the lens.

For the time delay $\Delta^2 t_\pm$, substituting Eq. \eqref{eq:r0solcase2inbeta} into \eqref{eq:totaltfirstfew}, and then further into \eqref{eq:tdpmdef} and simplify the result, we find
\begin{align}
\Delta^2 t_\pm=&\sum_{i=s,d} \frac{E}{ \sqrt{E^2- \kappa }} \frac{2 l^2 \sqrt{\eta +1} }{ r_i} + \frac{E l^2 \sqrt{\eta +1} }{6 \lb E^2-\kappa \rb^{\frac{3}{2} } } \frac{\Lambda}{ r_i} \nonumber \\
  & \times \lsb \lb 2 E^2 - 9 \kappa \rb l^2 \lb \eta +2\rb + 6 \kappa r_i^2 \rsb \nonumber \\
  & + \frac{E}{2 \lb E^2-\kappa \rb^{\frac{3}{2} } } \lcb \lsb \lb E^2-\kappa\rb c_1 - E^2 f_1 \rsb \frac{ 2l }{r_i} \right. \nonumber \\
  & \left. -f_1 \lb 2 E^2 - 3 \kappa \rb \ln \frac{\sqrt{\eta +1}+1}{\sqrt{\eta +1}-1} \rcb\nn\\
  &+\mathcal{O}\lb\frac{r_i}{b^2}, \frac{r_ir_0^2\Lambda}{b},r_i(r_0^2\Lambda)^2\rb \label{eq:tdpmres} 
\end{align}
where the auxiliary parameters $l$ and $\eta$ are 
\begin{align}
l=&\frac{r_d \beta}{2 \lsb 1+\kappa r_s r_d\Lambda /\lsb 6 \lb E^2-\kappa \rb \rsb\rsb}, \\
\eta =&-\frac{4f_1 r_s \lb 2E^2-\kappa\rb}{r_d \lb r_s+r_d\rb \beta^2 \lb E^2-\kappa\rb} \lsb 1+\frac{\kappa r_s r_d \Lambda}{6\lb E^2-\kappa\rb}\rsb.
\end{align}
The first, second and third terms in Eq. \eqref{eq:tdpmres} are due to the $\Delta t_{0,0},~\Delta t_{0,1}$ and $\Delta t_{1,0}$ orders in the total travel times \eqref{eq:totaltfirstfew} respectively. Therefore we can expect that the contribution from first term to the time delay should be larger than other terms. Note that in the case of $\Lambda=0$, Eq. \eqref{eq:tdpmres} agrees with the time delay in asymptotically flat SSS spacetimes \cite{Liu:2020wcu}. 
The dependence of \eqref{eq:tdpmres} on $\Lambda$ however is not only through the second term since there also exist \mclambda dependence in $l$ and $\eta$. To isolate the effect of \mclambda, we can further expand \eqref{eq:tdpmres} in the small \mclambda limit, and find to the leading order of \mclambda
\begin{align}
\Delta^2t_{\pm}\approx &\sum_{i=s,d} \frac{E}{ \sqrt{E^2- \kappa }} \lb \frac{2 l_0^2 \sqrt{\eta_0 +1} }{ r_i}  -f_1 \ln \frac{\sqrt{\eta_0 +1}+1}{\sqrt{\eta_0 +1}-1}  \rb\nn\\
&+ \frac{\Lambda E }{\lb E^2-\kappa\rb^{\frac{3}{2}}   } \lsb      \frac{-\kappa l_0^2r_s r_d \lb 3\eta_0+4 \rb}{ 6\sqrt{\eta_0+1}r_i} +\frac{\kappa  f_1 r_sr_d}{6\sqrt{\eta_0+1}}\right.\nn\\
&\left. + \sqrt{\eta_0+1}\lb \kappa  l_0^2  r_i 
  +\frac{l_0^4 \lb \eta_0 +2\rb (E^2-\kappa)}{3 r_i}\rb \rsb\nn\\
&+\mathcal{O}(\Lambda^2)\label{eq:tdpmlamexp}
\end{align}
where 
\begin{align}
l_0=&l(\Lambda\to0)=\frac{r_d \beta}{2},\\
\eta_0 =&\eta(\Lambda\to0)= - \frac{8f_1r_s}{r_d\lb r_s+r_d \rb\beta^2}.
\end{align}
The first, second and third terms of the $\mathcal{O}(\Lambda^1)$ order come respectively from the $\Delta t_{0,0}$ order, the $\Delta t_{1,0}$ order and $\Delta t_{0,1}$ order in the total travel time. 

Here it is important to note that in the time delay $\Delta^2 t_\pm$ in Eq. \eqref{eq:tdpmres}, three non-trivial spacetime parameters, $c_1,~f_1$ and $\Lambda$ are involved. However  the first term of the $\Delta t_{0,1}$ order given in the curl bracket, which contains the only occasion of $c_1$, can be shown to be always much smaller than the logarithmic term. And therefore in the expansion in Eq. \eqref{eq:tdpmlamexp}, the $c_1$ disappears in the leading order of $\Delta^2 t_\pm$. Therefore one can conclude that for arbitrary asymptotically dS spacetimes, the time delay $\Delta^2 t_\pm$ depends only on $f_1$ and $\Lambda$, to the leading order. 

\begin{figure}[htp!]
    \centering
    \includegraphics[width=0.45\textwidth]{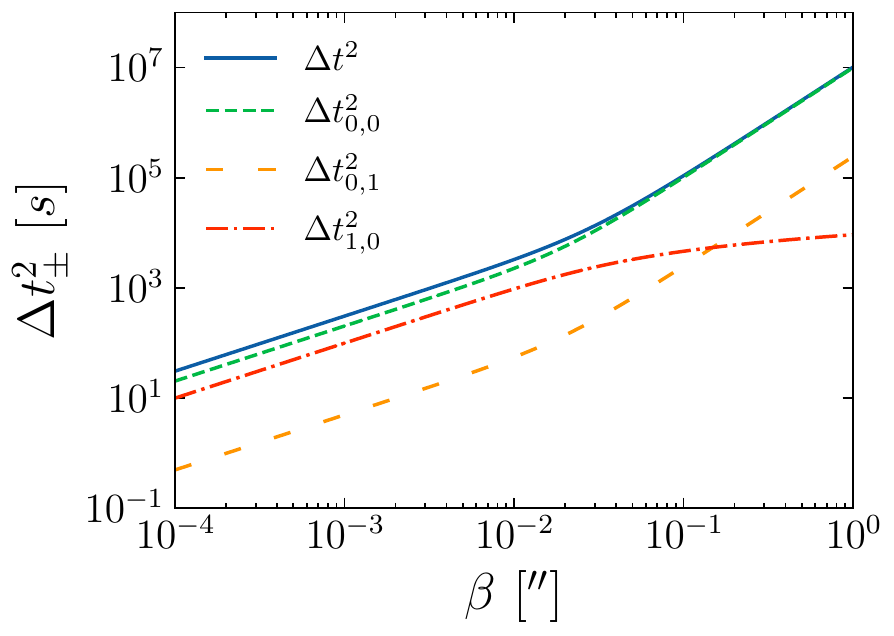}\\
    (a)\\
    \includegraphics[width=0.45\textwidth]{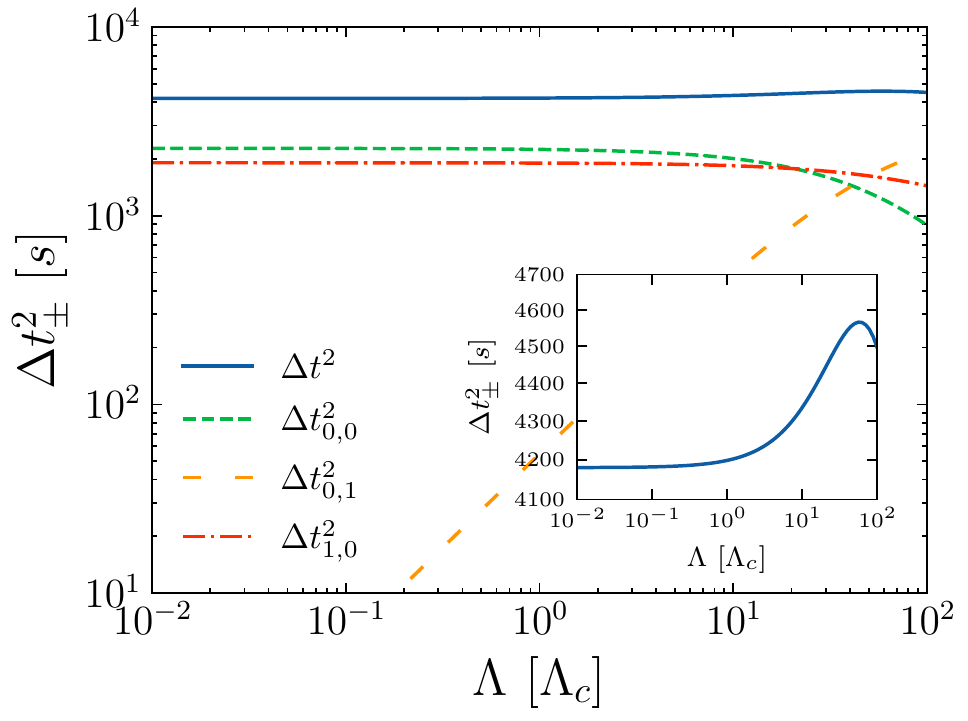}\\
    (b)
    \caption{Time delay $\Delta^2 t_\pm$ in Eq. \eqref{eq:tdpmres} and its various terms as a functions of $\beta$ for $\Lambda=\Lambda_c$ (a) and as functions of $\Lambda$ for $\beta=0.01~[^{\prime\prime}]$ (b). The inset shows the the total $\Delta ^2t_\pm$ using linear scale for y-axis. Other parameters are the same as in Fig. \ref{fig:r0thetaplot}. }
    \label{fig:dtpmplot}
\end{figure}

In Fig. \ref{fig:dtpmplot}, we plot the time delay \eqref{eq:tdpmres} and its three terms $\Delta^2 t_{\pm 0,0},~\Delta^2 t_{\pm0,1}$ and $\Delta^2 t_{\pm1,0}$ corresponding to  contributions from order $\Delta t_{0,0},~\Delta t_{0,1}$ and $\Delta t_{1,0}$ in the total travel time, as functions of $\beta$ and $\Lambda$. Note that for comparison purpose, in Fig. \ref{fig:dtpmplot} (b), we have allowed $\Lambda$ to go beyond $\Lambda_c$ to about $100\Lambda_c$. 
We observe from Fig. \ref{fig:dtpmplot} (a) that as \mbeta decreases to zero, all contributions decrease monotonically to zero, as expected for the time delay between signals from the two sides of the lens. When \mbeta is smaller than $0.007~[^{\prime\prime}]\sim \theta_\mathrm{E}$, the $\Delta^2t_{\pm0,0}$ and $\Delta^2t_{\pm1,0}$ contribute similarly. When \mbeta increases beyond $\theta_\mathrm{E}$ however,  the $\Delta^2t_{\pm0,1}$ increases much faster than $\Delta^2t_{\pm1,0}$ and surpasses it at about $\beta\approx 0.23~[^{\prime\prime}]$.
On the other hand,  from Fig. \ref{fig:dtpmplot} (b) we observe that as \mclambda increases to $\sim 10\Lambda_c$,  the $\Delta^2t_{\pm0,1}$ term increases almost linearly as dictated by the second term of Eq. \eqref{eq:tdpmres} and its contribution surpasses the $\Delta^2t_{\pm1,0}$ and $\Delta^2t_{\pm0,0}$ terms when $\Lambda=50.6\Lambda_c$ and $\Lambda=41.1\Lambda_c$ respectively. 
Moreover, we also note that for $\Lambda$ smaller than $10\Lambda_c$,
superficially both the $\Delta^2t_{\pm0,0}$ and $\Delta^2t_{\pm1,0}$ terms are almost invariant as \mclambda changes. However, from Eq. \eqref{eq:tdpmlamexp} it is known that indeed these two terms also contain \mclambda dependence comparable to the $\Delta^2t_{\pm0,1}$ term in at least some parameter space spanned by $\beta,~\Lambda$ and $E$ etc. Indeed, as $\Lambda$ increases larger than $10\Lambda_c$, their dependence on $\Lambda$ becomes very apparent. Furthermore, in the three terms proportional to $\Lambda$ in Eq. \eqref{eq:tdpmlamexp}, we can show that for any reasonable set of parameters $\{r_d,~M,~\beta\}$, the third term will always be positive and larger than the first two terms and therefore the total $\Delta^2t_\pm$ should increase as the small $\Lambda$ increase. This is also seen in the inset of Fig. \ref{fig:dtpmplot} (b), in which $\Delta^2 t_\pm$ increased until $\Lambda\approx 59\Lambda_c$, beyond which the small $\Lambda$ expansion \eqref{eq:tdpmlamexp} broke down.  

For the time delay between signals from the same side, substituting $r_{0\pm}$ in Eq. \eqref{eq:r0solcase2inbeta} with the same sign but different energies $E_1$ and $E_2$ into \eqref{eq:totaltfirstfew}, and then further into \eqref{eq:td12def}, the value of  $\Delta^2 t_{12}$ can be obtained. Since the signals we consider are usually relativistic or null, we can actually substitute \eqref{eq:r0solcase2inbeta} directly into \eqref{eq:tdlargee} and then expanding it in the large energy limit and directly find, to the leading order of $1/E^2$,   \begin{align}
\Delta^2 t_{12}=&\sum_{i=s,d}  \lb \frac{1}{E_1^2} - \frac{1}{E_2^2} \rb \lsb \frac{r_i}{2}  - \frac{r_i r_{0\xi \infty}^2   \Lambda}{4} \rsb + \mathcal{O} \lb \frac{r_i}{E^4} \rb
\label{eq:td12sim}
\end{align}
where
\begin{align}
r_{0\xi \infty} \equiv&  r_{0\xi }\lb E\to\infty \rb \nn\\
=& l_0 \lb \sqrt{\eta_0+1}-\xi \rb,~\xi =\pm1 
\end{align}
It is seen that generally, the existence of a positive cosmological constant will decrease the time delay $\Delta^2t_{12}$.  
The second and first terms in the square bracket of Eq. \eqref{eq:td12sim} have a ratio of $r_{0\xi \infty}^2\Lambda/2$. 
Unfortunately, for the currently known $\Lambda_c$, $r_{0\xi\infty}^2\Lambda_c\ll 1$ and therefore the second term involving $\Lambda$ will be much smaller than the first term. 
Let us take the time delay between different mass eigenstates $|\nu_i\rangle ~(i=1,2,3)$ of supernova neutrinos as an example.
Assuming that the average energy of the mass eigenstates is $\langle E\rangle=10~[\text{MeV}]$, then the energies per unit mass of the first mass eigenstates $|\nu_j\rangle$ and the second mass eigenstate $|\nu_k\rangle$ are respectively $E_1=\langle E\rangle /m_j$ and $E_2=\langle E\rangle /m_k$. Substituting them into Eq. \eqref{eq:td12sim}, then the time delay between them simplifies to
\begin{align}
\Delta^2 t_{12}=&\sum_{i=s,d}   \frac{\Delta m^2_{jk}}{\langle E\rangle^2}\lsb \frac{r_i}{2}  - \frac{r_i r_{0\xi \infty}^2   \Lambda}{4} \rsb + \mathcal{O} \lb \frac{r_i}{E^4} \rb
\label{eq:td12simnet}
\end{align}
Since the mass square differences of neutrino mass eigenstates are  \cite{ParticleDataGroup:2020ssz}
\begin{align}
\Delta m^2_{12}=-7.53\times 10^{-5} ~[\text{eV}^2],
\end{align} 
and for normal ordering 
\begin{align}\Delta m^2_{23}=-2.453 \times 10^{-3} [\text{eV}^2]\end{align}
and inverse ordering 
\begin{align} \Delta m^2_{23}=2.546 \times 10^{-3} ~[\text{eV}^2],
\end{align}
then for the time delay between the mass eigenstates $|\nu_1\rangle$ (or roughly $|\nu_2\rangle$) and $|\nu_3\rangle$, one can compute the first and second terms in the square bracket of Eq. \eqref{eq:td12simnet} using the same $r_{s,d}$ and $M$ as in Figs. \ref{fig:r0thetaplot}. They yield at most $10.7$ [s] and $2.14\times 10^{-8}$ [s] for $\beta\approx 10~[^{\prime\prime}]$ and $\xi=-1$. 
Only if $r_{0\xi\infty}^2\Lambda $ were larger, which requires a larger $r_{s,d},~M$ and $\beta$, or the resolution of the time delay is much better than $\mathcal{O}(10^{-8})$ [s], will the effect of $\Lambda$ on $\Delta^2t_{12}$ be explicitly distinguishable. 

\section{Conclusion\label{sec:conc}}

In this paper, we studied the deflection and GL of null and timelike rays in general SSS and asymptotically (a)dS spacetimes. 
We first showed that if the metric function satisfies a simple condition, Eq. \eqref{eq:metriccform}, then the change of the angular coordinate for fixed $r_0$ will actually not depend on the parameter \mclambda for null rays, although that of timelike rays \eqref{eq:dphitimelike}, the apparent angles of the GL images \eqref{eq:aasol}, and the total travel time \eqref{eq:totaltsub} will still depend on \mclambda explicitly. 
A two-step perturbative method is then developed to calculate the change of the angular coordinate and the result is expressed into a quasi-series form of both $M/r_0$ and $\Lambda$, i.e. Eq. \eqref{eq:alphaintres}, with the coefficients of the series determined by the asymptotic expansion coefficients of the metric functions. The finite distance effect of the source/detector is also naturally taken into account. The methodology is applied to SdS, RNdS, dilaton-dS and brane world-dS spacetimes to find the deflection in them. The total travel time is computed using the same method and expressed into a similar quasi-series, Eq. \eqref{eq:tintres}. 

Using the obtained deflection, an exact GL equation is used to solve the allowed minimal radius \eqref{eq:r0solcase2inbeta} from each side of the lens and the corresponding apparent angles. It is found that generally, due to the smallness of currently known \mclambda, only for source of very large distance can the effect of \mclambda be apparent on these quantities. In this case, as \mclambda increases, both apparent angles will decrease as $\Lambda$ increases (see Fig. \ref{fig:r0thetaplot}). The time delays $\Delta^2t_\pm$ between the images from two sides and $\Delta^2 t_{12}$ between signals from the same side are also obtained in Eqs. \eqref{eq:tdpmres} and \eqref{eq:td12sim}. We found that qualitatively a small positive $\Lambda$ will cause the former to increase while the latter to decrease. 

We emphasize the generality of the perturbative method developed in this work: besides the four spacetimes considered in Sec. \ref{sec:spacetimecases}, it can be applied to other interesting asymptotically dS spacetimes for the computation of both deflection and time delay for both null and timelike rays, with finite distance effect naturally taken into account. Moreover, although in the examples in this work, we explicitly consider the case of asymptotically de Sitter spacetimes with $\Lambda>0$, indeed both the method and relevant results are valid for asymptotically anti-de Sitter spacetimes. Finally, 
a generalization of the method to more realistic spacetime such as the equatorial plane of Kerr-dS or Kerr-Newmann-dS spacetimes is also possible.

\acknowledgements
The authors thank Ke Huang, Chen Huang and Tingyuan Jiang for valuable discussions. This work is supported by the NNSF China 11504276 and MOST China 2014GB109004.

\appendix

\section{Integration formulas \label{sec:appd}}
To carry out the integrals in Eq. \eqref{eq:ttintfuncu}, we first need to do partial fraction decomposition to the integrand. 
When $k+1< 0$, we have
\begin{widetext}
\begin{align}
\frac{\cos^{n-k-1}\alpha }{\lb \cos\alpha +1 \rb^n} =& \sum_{i=0}^{n} \frac{C_{n-k-1}^i (-1)^{n-k-i-1}}{\lb 1+\cos \alpha\rb^{n-i}} + \sum_{i=1}^{-k-1} \sum_{j=0}^{i} C_{n-k-1}^{i+n} C_{i}^{j} (-1)^{k+i+1} \cos^j \alpha, \nn\\
 = & \sum_{i=0}^{n} \frac{C_{n-k-1}^i (-1)^{n-k-i-1}}{\lb 2 \cos^2 \frac{\alpha}{2}\rb^{n-i}} + \sum_{i=1}^{-k-1} \sum_{j=0}^{i} C_{n-k-1}^{i+n} C_{i}^{j} (-1)^{k+i+1}  \cos^j \alpha.
\end{align}
When $0\leq k+1\leq n$,  we have
\begin{align}
\frac{\cos^{n-k-1}\alpha }{\lb \cos\alpha +1 \rb^n} = \sum_{i=0}^{n-k-1} \frac{C_{n-k-1}^i (-1)^{n-k-i-1}}{\lb 2 \cos^2 \frac{\alpha}{2}\rb^{n-i}}.
\end{align}
When $n<k+1$, we have
\begin{align}
\frac{\cos^{n-k-1}\alpha }{\lb \cos\alpha +1 \rb^n} =& (-1)^{k-n+1}\sum_{i=1}^{n} \frac{C_{k-i}^{n-i}}{\lb 2 \cos^2 \frac{\alpha}{2}\rb^{i}} + \sum_{j=1}^{k-n+1} \frac{(-1)^{k-n-j+1}C_{k-j}^{k-n-j+1}}{\cos^j \alpha}. 
\end{align}
Eq. \refer{eq:ttintfuncu} is integrable if the terms on the right hand sides of these equations are integrable. Their integration can be done using Eqs. $\mathbf{2.513}\ 3,4$ and $\mathbf{2.519}\ 1,2$ in Ref. \cite{Gradshteyn:2007} and the results for $l>0$ are
\begin{subequations}\label{eqs:intcosalpha}
\begin{align}
\int \frac{\dd \alpha}{\cos^{2l} \alpha} =& \frac{\sin \alpha}{2 l - 1}\lsb \sec^{2l-1} \alpha + \sum_{k_l=1}^{l-1} \frac{2^{k_l}\lb l-1\rb(l-2)\dots(l-k_l)}{(2l-3)(2l-5)\dots (2l-2k_l -1)} \sec^{2l-2k_l -1}\alpha \rsb, \label{eqs:intcosalphaa} \\
\int \frac{\dd \alpha}{\cos^{2l+1} \alpha} =& \frac{\sin \alpha}{2l} \lsb \sec^{2l} \alpha + \sum_{k_l=1}^{l-1} \frac{(2l-1)(2l-3)\dots (2l-2k_l +1)}{2^{k_l}\lb l-1\rb(l-2)\dots(l-k_l)} \sec^{2l-2k_l}\alpha \rsb \nonumber \\
 & + \frac{(2l-1)!!}{2^l l!} \ln \tan\lb \frac{\pi}{4} + \frac{\alpha}{2} \rb, \label{eqs:intcosalphab} \\
\int \cos^{2l} \alpha \dd \alpha =& \frac{1}{2^{2l} } C_{2l}^{l} \alpha + \frac{1}{2^{2l-1}} \sum_{k_l=0}^{l-1} C_{2l}^{k_l} \frac{\sin \lsb (2l-2k_l) \alpha \rsb}{2l-2k_l},\label{eqs:intcosalphac}\\
\int \cos^{2l+1} \alpha \dd \alpha = & \frac{1}{2^{2l}} \sum_{k_l=0}^{l} C_{2l+1}^{k_l} \frac{\sin \lsb (2l-2k_l+1) \alpha \rsb}{2l-2k_l+1}.
\label{eqs:intcosalphad}
\end{align}
\end{subequations}
Corresponding to Eqs. \eqref{eqs:intcosalphaa} and \eqref{eqs:intcosalphab}, and \eqref{eqs:intcosalphac} and \eqref{eqs:intcosalphad}, we then define two functions $F_n(\alpha)$ and $G_n(\alpha)$ as
\begin{subequations}\label{eqs:intalphatot}
\begin{align}
\mathrm{F}_0(\alpha)=& \alpha \\
\mathrm{F}_1(\alpha)=& \ln \tan\lb \frac{\pi}{4} + \frac{\alpha}{2} \rb \\
\mathrm{F}_n(\alpha)=& \frac{\sin \alpha}{n - 1}\lsb \sec^{n-1} \alpha + \sum_{i=1}^{\lsb \frac{n}{2} \rsb -1} \frac{\lb n-2\rb(n-4)\dots(n-2i)}{(n-3)(n-5)\dots (n-2i -1)} \sec^{n-2i -1}\alpha \rsb \nonumber \\
  & + \lcb \begin{array}{ll} 0,\ \ &\text{$n$ is even and $n>0$,} \\
 \frac{(n-2)!!}{(n-1)!!} \ln \tan\lb \frac{\pi}{4} + \frac{\alpha}{2} \rb, \ \ &\text{$n$ is odd and $n>1$,}  \end{array} \right. \\
\mathrm{G}_0(\alpha)=& \alpha \\
\mathrm{G}_1(\alpha)=& \sin \alpha \\
\mathrm{G}_n(\alpha)=& \frac{1}{2^{n-1}} \sum_{i=0}^{\lsb \frac{n+1}{2} \rsb -1} C_{n}^{i} \frac{\sin \lsb (n-2i) \alpha \rsb}{n-2i} + \lcb \begin{array}{ll} \frac{1}{2^n } C_{n}^{\frac{n}{2}} \alpha,\ \ &\text{$n$ is even and $n>0$,}\\ 0, \ \ &\text{$n$ is odd and $n>1$,}  \end{array} \right.
\end{align}
\end{subequations}
So the result of the integral Eq. \refer{eq:ttintfuncu} is
\bea 
I_{n,k}
&=&\int_{0}^{\alpha_j} \frac{\cos^{n-k-1}\alpha }{\lb \cos\alpha +1 \rb^n} \dd\alpha~~~(\alpha_j=\sec^{-1}u_j,~j=s,d) \nn\\
&=& \lcb \begin{array}{ll}  \displaystyle \sum_{i=0}^{n} \frac{C_{n-k-1}^i (-1)^{n-k-i-1}}{2^{n-i-1}}\mathrm{F}_{2n-2i}\lb \frac{\alpha_j}{2} \rb  + \sum_{i=1}^{-k-1} \sum_{j=0}^{i} C_{n-k-1}^{n+i} C_{i}^{j} (-1)^{k+i+1} \mathrm{G}_j\lb \alpha_j\rb,  & ~~k+1<0, \\
\displaystyle \sum_{i=0}^{n-k-1} \frac{C_{n-k-1}^i (-1)^{n-k-i-1}}{2^{n-i-1}}\mathrm{F}_{2n-2i}\lb \frac{\alpha_j}{2} \rb, & ~~0\leq k+1< n, \\
\displaystyle  (-1)^{k-n+1}\sum_{i=1}^{n} \frac{C_{k-i}^{n-i}}{2^{i-1}} \mathrm{F}_{2i}\lb \frac{\alpha_j}{2} \rb  + \sum_{i=1}^{k-n+1} (-1)^{k-n-i+1}C_{k-i}^{k-n-i+1} \mathrm{F}_i\lb \alpha_j \rb,  & ~~n<k+1. \end{array} \right.
\label{eq:inkres}\eea
\end{widetext}

When $n,~k$ are small, the first few $I_{n,k}$ that will be used in the main text can be given explicitly
\begin{subequations}
\label{eqs:inks}
\begin{align}
%I_{0,-1}(u_j)=&\frac{\pi}{2}-\tan ^{-1}\left(\frac{1}{\sqrt{u_j^2-1}}\right), \\
%I_{0,0}(u_j)=&\ln \left(\sqrt{u_j^2-1}+u_j\right), \\
I_{0,1}(u_j)=&\sqrt{u_j^2-1}, \\
I_{0,2}(u_j)=& \frac{u_j \sqrt{u_j^2-1}}{2}+\frac{1}{2}\ln \left(\sqrt{u_j^2-1}+u_j\right), \\
I_{0,3}(u_j)=&\frac{1}{3} \sqrt{u_j^2-1} \left(u_j^2+2\right), \\
%%I_{0,4}(u_j)=&\frac{1}{8} \left(u_j \left(2 u_j^2+3\right) \sqrt{u_j^2-1}+3 \ln \left(\sqrt{u_j^2-1}+u_j\right)\right), \\
%I_{1,-2}(u_j)=& \left(\frac{1}{u_j+1}+\frac{1}{u_j}\right) \sqrt{u_j^2-1}+\tan ^{-1}\left(\frac{1}{\sqrt{u_j^2-1}}\right) - \frac{\pi}{2}, \\
%I_{1,-1}(u_j)=& \frac{\pi}{2} -\frac{\sqrt{u_j^2-1}}{u_j+1}-\tan ^{-1}\left(\frac{1}{\sqrt{u_j^2-1}}\right), \\
I_{1,0}(u_j)=&\frac{u_j-1}{\sqrt{u_j^2-1}}, \\
I_{1,1}(u_j)=&\ln \left(\sqrt{u_j^2-1}+u_j\right)-\frac{\sqrt{u_j^2-1}}{u_j+1}.
%I_{1,2}(u_j)=&\frac{u_j^2+u_j-2}{\sqrt{u_j^2-1}} - 2 \ln \frac{\sqrt{u_j-1}+\sqrt{u_j+1}}{\sqrt{2}} , \\
%I_{1,3}(u_j)=&\frac{u_j^3-2 u_j^2-3 u_j+4}{2 \sqrt{u_j^2-1}}+3 \ln \frac{\sqrt{u_j-1}+\sqrt{u_j+1}}{\sqrt{2}}.
\end{align}
\end{subequations}

The large $u_j$ expansion of $I_{n,k}(u_j)$ can also be worked out to the first non-trivial order as
\begin{align} I_{n,k}\approx \begin{cases} 
\displaystyle \frac{u_{s,d}^{k-n}}{k-n}= \frac1{k-n}\lb\frac{r_{s,d}}{r_0}\rb^{k-n},&~ k> n\\
\\
\ln(2u_j)=\ln\lsb \frac{2r_{s,d}}{r_0}\rsb ,&~k=n,\\
\\
\displaystyle \frac{(n-k-1)!}{2^k(2n-2k-1)!!}\\
~~\times~_2F_1\lb \frac12,-k;\frac12+n-k;-1\rb\equiv L_{n,k} ,&~k< n.\end{cases}
\label{eq:inkgeneralexp}
\end{align}
where $_2F_1$ is the Gaussian hypergeometric function and for notation simplicity, we have denoted the result for the case $k<n$ as $L_{n,k}$. 
For the fist few $n$ and $k$ that will be used, their expansions can be worked to even higher orders,
\begin{subequations}
\label{eq:i1nexp}
\begin{align}
I_{0,-1}=&\frac{\pi}{2}-\frac{1}{u_j} + \mathcal{O}\lb \frac{1}{u_j^2} \rb, \\
I_{0,0}=& \ln( 2u_j) + \mathcal{O}\lb \frac{1}{u_j^2} \rb, \\
I_{0,1}=& u_j - \frac{1}{2u_j} + \mathcal{O}\lb \frac{1}{u_j^2} \rb , \\
I_{0,2}=& \frac{u_j^2}{2} + \frac{\ln 2 u_j}{2} - \frac{1}{4} + \mathcal{O}\lb \frac{1}{u_j^2} \rb , \\
I_{0,3}=&\frac{1}{3} \sqrt{u_j^2-1} \left(u_j^2+2\right), \\
%I_{0,4}=&\frac{1}{8} \lsb u_j \left(2 u_j^2+3\right) \sqrt{u_j^2-1}+3 \ln \left(\sqrt{u_j^2-1}+u_j\right)\rsb, \\
%I_{0,5}=&\frac{1}{15} \sqrt{u_j^2-1} \left(3 u_j^4+4 u_j^2+8\right), \\
I_{1,-2}=& 2 - \frac{\pi}{2} + \mathcal{O}\lb \frac{1}{u_j^2} \rb, \\
I_{1,-1}=& \frac{\pi}{2} - 1  + \mathcal{O}\lb \frac{1}{u_j^2} \rb, \\
I_{1,0}=& 1 - \frac{1}{u_j} + \mathcal{O}\lb \frac{1}{u_j^2} \rb , \\
I_{1,1}=& \ln(2u_j) - 1 + \frac{1}{u_j} + \mathcal{O}\lb \frac{1}{u_j^2} \rb, \\
I_{1,2}=& u_j - \ln(2u_j) + 1 - \frac{3}{2u_j} + \mathcal{O}\lb \frac{1}{u_j^2} \rb , \\
%I_{1,3}=&\frac{u_j^3-2 u_j^2-3 u_j+4}{2 \sqrt{u_j^2-1}}+3 \ln \frac{\sqrt{u_j-1}+\sqrt{u_j+1}}{\sqrt{2}}, \\
%I_{1,4}=&\frac{2 u_j^4-3 u_j^3+8 u_j^2+9 u_j-16}{6 \sqrt{u_j^2-1}} - 3 \ln \left(\sqrt{u_j-1}+\sqrt{u_j+1}\right), \\
I_{2,-3}=& \frac{7\pi}{4} -\frac{16}{3} + \mathcal{O}\lb \frac{1}{u_j^2} \rb , \\
I_{2,-2}=& \frac{10}{3} - \pi + \mathcal{O}\lb \frac{1}{u_j^2} \rb, \\
I_{2,-1}=& \frac{\pi}{2} - \frac{4}{3} + \mathcal{O}\lb \frac{1}{u_j^2} \rb , \\
I_{2,0}=& \frac{1}{3} + \mathcal{O}\lb \frac{1}{u_j^2} \rb , \\
I_{2,1}=& \frac{2}{3} - \frac{1}{u_j} + \mathcal{O}\lb \frac{1}{u_j^2} \rb .
%I_{2,2}=& \ln (2u_j) - \frac{5}{3} + \frac{2}{u_j} + \mathcal{O}\lb \frac{1}{u_j^2} \rb . 
%I_{2,3}=&\frac{3 u_j^3+11 u_j^2-4 u_j-10}{3 (u_j+1) \sqrt{u_j^2-1}} - 4 \ln \left(\sqrt{u_j-1}+\sqrt{u_j+1}\right).
\end{align}
\end{subequations}

\end{document}